\documentclass[11pt]{article}
\pdfoutput=1

\usepackage{hyperref}
\usepackage[normalem]{ulem} 
\usepackage{cite}
\usepackage{appendix}
\usepackage{epsfig}
\usepackage{amsmath}
\usepackage{amssymb}
\usepackage{bm}
\usepackage{hyperref}
\usepackage{slashed}

\newcommand{\be} {\begin{equation}}
\newcommand{\ee} {\end{equation}}
\newcommand{\ba} {\begin{eqnarray}}
\newcommand{\ea} {\end{eqnarray}}

\newcommand{\no} {\nonumber}
\newcommand{\cL} {\mathcal L}
\newcommand{\cA} {\mathcal A}
\newcommand{\cB} {\mathcal B}

\newcommand{\cLT} {{ \mathcal L}^{T}}
\newcommand{\cLS} {{ \mathcal L}^{S}}
\newcommand{\llq} {\lambda^q}
\newcommand{\lld} {\lambda^d}
\newcommand{\lle} {\lambda^\ell}

\newcommand{\RK} {R_{K^{(*)}}^{\mu e}}
\newcommand{\RD} {R_{D^{(*)}}^{\tau \ell}}

\usepackage{color}
\definecolor{darkblue}{cmyk}{1,0.3,0,0.2}
\definecolor{violet}{cmyk}{0,1,0,0.2}
\hypersetup{colorlinks, bookmarksnumbered, citecolor=darkblue, linkcolor=darkblue, pdfstartview=FitH, urlcolor=darkblue, linktocpage}

\begin{document}

\begin{flushright}
 ZU-TH-18/17  \\
\end{flushright}

\thispagestyle{empty}

\bigskip

\begin{center}
\vspace{1.5cm}
     {\Large\bf $\boldsymbol{B}$-physics anomalies: a guide to combined explanations} \\ [1cm] 

   {\bf Dario Buttazzo$^{a}$, Admir Greljo$^{a,b}$, Gino Isidori$^{a}$, David Marzocca$^a$}    \\[0.5cm]
  {\em $(a)$  Physik-Institut, Universit\"at Z\"urich, CH-8057 Z\"urich, Switzerland}  \\
  {\em $(b)$  Faculty of Science, University of Sarajevo, Zmaja od Bosne 33-35, \\ 71000 Sarajevo, Bosnia and Herzegovina}  \\  
\end{center}
\vspace{1cm}

\centerline{\large\bf Abstract}
\begin{quote}
Motivated by additional experimental hints of Lepton Flavour Universality violation in $B$ decays, 
both in charged- and in neutral-current processes, we analyse the ingredients necessary to provide a 
combined description of these phenomena.  By means of an Effective Field Theory (EFT) approach,
based on the hypothesis of New Physics coupled predominantly to the  third generation 
of left-handed quarks and leptons, we show how this is possible.
We demonstrate, in particular, how to solve the problems posed by electroweak precision tests and direct searches 
with a rather natural choice of model parameters, within the context of a $U(2)_q \times U(2)_\ell$ flavour symmetry.  
We further exemplify the general EFT findings by means of simplified models with explicit mediators in the TeV range: 
coloured scalar or vector leptoquarks and colour-less vectors. 
Among these, the case of an $SU(2)_L$-singlet vector leptoquark emerges 
as a particularly simple and  successful framework.
\end{quote}

\newpage
\tableofcontents
\newpage

\section{Introduction}

One of the most interesting phenomena reported by particle physics 
experiments in the last few years are the numerous hints of Lepton Flavour Universality (LFU) 
violations observed in semi-leptonic $B$ decays.
The very recent LHCb results on the LFU ratios $\RK$~\cite{Aaij:2017vbb}
and $\RD$~\cite{LHCbRDstar}
are the last two pieces of a seemingly coherent set of anomalies  
which involves different observables 
and 
experiments.
So far, not a single LFU ratio measurement
exhibits
a deviation with respect to the 
Standard Model (SM) above the $3\sigma$ level. 
However, the overall set of observables 
is very consistent and, once combined, the probability of a 
mere statistical fluctuation is very low.

The evidences collected so far can naturally be grouped into two categories,
according to the underlying quark-level transition:
\begin{itemize}
\item deviations from 
$\tau/\mu$ (and $\tau/e$) universality in  $b\to c \ell \bar\nu$
charged currents~\cite{Lees:2013uzd,Hirose:2016wfn,Aaij:2015yra,LHCbRDstar};
\item deviations from 
$\mu/e$ universality in  $b\to s \ell \overline{\ell}$ neutral currents~\cite{Aaij:2014ora,Aaij:2017vbb}.
\end{itemize}
In both cases the combination of the results leads to an evidence around 
the $4\sigma$ level for LFU {violating} contributions of non-SM origin, 
whose size is  
$\mathcal{O}(10\%)$ 
compared to the corresponding 
charged- or neutral-current SM amplitudes.
Furthermore, a strong evidence for a deviation from the SM prediction has been observed by LHCb in the angular distribution
of the $B^0 \to K^{*0} \mu^+ \mu^-$ decay \cite{Aaij:2013qta,Aaij:2015oid}, which is consistent with the deviations from LFU in neutral-current $B$ decays~\cite{Altmannshofer:2015sma,Descotes-Genon:2015uva}.

These deviations from the SM have triggered a series of theoretical speculations
about possible New Physics (NP) interpretations. Attempts to provide 
a combined/coherent explanation for  both charged- and neutral-current anomalies have been
presented  in Refs.~\cite{Bhattacharya:2014wla, Alonso:2015sja, Greljo:2015mma, Calibbi:2015kma,Bauer:2015knc,
Fajfer:2015ycq, Barbieri:2015yvd, Das:2016vkr, Boucenna:2016qad,Becirevic:2016yqi,  Hiller:2016kry, Bhattacharya:2016mcc,
Buttazzo:2016kid,Barbieri:2016las,Bordone:2017anc,Crivellin:2017zlb,Becirevic:2016oho,Cai:2017wry,Megias:2017ove}.
A  common origin of the two 
set of anomalies is not obvious, but is very appealing since: i)~in both  types of semi-leptonic $B$-meson decays  (charged and neutral)
we are dealing with a violation of LFU;
ii)~in both cases data favours left-handed effective interactions that, due to the SM gauge symmetry, naturally suggest a connection between charged and neutral currents.

One of the puzzling aspects of the present anomalies is that they have been observed only in semi-leptonic $B$ decays and are quite large compared to the corresponding SM  amplitudes. 
On the contrary, no evidence of deviation from the SM has been seen so far in the precise (per-mil) tests of LFU in semi-leptonic $K$ and $\pi$ decays, purely leptonic $\tau$ decays, and in the electroweak precision observables. 
The most natural assumption to address this apparent paradox is the hypothesis that the NP responsible for the breaking of LFU is coupled mainly to the third generation of quarks and leptons, with a small (but non-negligible) mixing with the light generations~\cite{Greljo:2015mma,Bordone:2017anc,Glashow:2014iga}.
This hypothesis also provides a natural first-order explanation for the different size 
of the two effects, which compete with a tree-level SM amplitude in charged currents,
and with a suppressed loop-induced SM amplitude in neutral currents, respectively.  
Within this paradigm, a class of particularly motivated models 
includes those which are based   
on a $U(2)_q\times U(2)_\ell$  flavour symmetry acting 
on the light generations of SM fermions~\cite{Barbieri:2011ci,Barbieri:2012uh},
and new massive bosonic mediators around the TeV scale:
colour-less vector $SU(2)_L$-triplets ($W^\prime$, $B^\prime$)~\cite{Greljo:2015mma},
vector $SU(2)_L$-singlet or -triplet leptoquarks (LQ)~\cite{Barbieri:2015yvd}, 
or scalar $SU(2)_L$-singlet and -triplet leptoquarks.
Besides providing a good description of low-energy data, 
these mediators could find a consistent UV completion in the context of 
strongly-interacting theories with new degrees of freedom at the TeV 
scale~\cite{Buttazzo:2016kid,Barbieri:2016las}.  
 
While these NP interpretations are quite interesting, their compatibility with the 
high-$p_T$  data from the LHC and other low-energy precision observables
is not trivial. On the one hand,  
high-$p_T$ searches for resonances (colour-less vectors in $s$-channel) or smooth distortions (leptoquarks in $t$-channel) in the $\tau\bar\tau$ invariant mass distribution ($pp \to\tau\bar\tau + X$) 
put very stringent constraints on a large class of 
models addressing the $\RD$ anomalies~\cite{Faroughy:2016osc}.
On the other hand, the  consistency with precise data on $\tau$ leptonic decays 
and  $Z$-boson  effective couplings, 
after taking into account quantum corrections, seems to be problematic~\cite{Feruglio:2016gvd,Feruglio:2017rjo}.
Last but not least, in most
explicit models constructed so far, a non-negligible 
amount of  fine-tuning  
is unavoidable in order 
to satisfy the constraints from $B_s$ and $B_d$ meson-antimeson 
mixing (see, in particular, Refs.~\cite{Barbieri:2015yvd,Buttazzo:2016kid}).

Motivated by the increased statistical significance of both 
sets of anomalies~\cite{Aaij:2017vbb,LHCbRDstar},  and  
focused on  finding a  common
explanation of the two effects within the same framework, 
in this paper we present a
combined analysis of these non-standard phenomena, 
addressing in detail the compatibility with all available low-energy observables,
electroweak precision tests, and high-$p_T$ searches. 
Updating, and significantly extending,  the first attempt of this type presented in 
Ref.~\cite{Greljo:2015mma}, we follow a bottom-up approach based on two main steps: 
\begin{enumerate}
\item[1.]
general EFT-type analysis of four-fermion semi-leptonic operators 
(addressing both semi-leptonic observables and radiatively induced effects in 
non-semi-leptonic processes), covering at the same time the 
underlying hypothesis of colour-less or LQ mediators;
\item[2.] exploration of the connections to 
other flavour and high-$p_T$ observables using simplified 
dynamical models for the possible sets of mediators. 
\end{enumerate}
In both cases we assume a minimally broken 
$U(2)_q\times U(2)_\ell$  flavour symmetry in order to constrain the 
flavour structure of the theory.

The paper is organised as follows. In Section~\ref{sec:EFT} we focus on the first step outlined above.
More explicitly, we analyse the flavour structure of the minimal set of semi-leptonic operators addressing the anomalies; we perform  
a fit of the Wilson coefficients of these operators to all the relevant semi-leptonic and purely leptonic (loop-induced) observables;
we discuss the interplay with the high-$p_T$ and $\Delta F =2$ processes based on the pure EFT-type considerations.
In Section~\ref{sect:models} we exemplify the findings of the previous section proposing three concrete (simplified) models which 
can simultaneously explain both anomalies while satisfying all available constraints from  low- and high-energy data.  
Finally, in Section~\ref{sec:UV} we briefly present some considerations about possible UV completions for the simplified models 
considered in Section~\ref{sect:models}.  
The results of our analysis are summarised in the Conclusions. Technical details concerning the flavour structure of the EFT 
and the observables entering the fit are presented in the Appendix.

\section{Semi-leptonic effective operators}
\label{sec:EFT}

In this section we analyse the flavour structure and the constraints on the semi-leptonic four-fermion operators 
contributing at the tree-level to  $\RK$ and $\RD$, taking into account the bounds from processes affected by the same effective operators both at the tree-level and beyond. 
We do not attempt a completely model-independent  EFT-type analysis, 
but we keep the discussion sufficiently general under the main hypothesis of NP coupled predominantly 
to third-generation left-handed quarks and leptons. 

More explicitly, our working hypotheses to determine the initial conditions of the EFT, at a scale $\Lambda$ 
above the electroweak scale, are the following:
\begin{itemize}
\item[1.] only four-fermion operators built in terms of left-handed quarks and leptons have non-vanishing Wilson coefficients;
\item[2.] the flavour structure is determined by the $U(2)_q\times U(2)_\ell$  flavour symmetry, minimally broken by two 
spurions $V_q \sim ({\bf 2},{\bf 1})$ and $V_\ell \sim ({\bf 1},{\bf 2})$; 
\item[3.] operators containing flavour-blind contractions 
of the light fields have vanishing Wilson coefficients.
\end{itemize}
We first discuss the consequences of these hypotheses on the structure of the relevant effective  
operators and then proceed analysing the experimental constraints on their couplings.

\subsection{The effective Lagrangian}

According to the first hypothesis listed above, we consider the following effective Lagrangian at a scale $\Lambda$ above the electroweak scale
\be
\mathcal{L}_{\rm eff} = \cL_{\rm SM}  - \frac{1}{v^2} \llq_{ij} \lle_{\alpha\beta} \left[C_{T}~(\bar Q_L^i \gamma_\mu \sigma^a Q_L^j ) (\bar L_L^\alpha \gamma^\mu \sigma^a L_L^\beta) +C_{S}~ (\bar Q_L^i \gamma_\mu Q_L^j) (\bar L_L^\alpha \gamma^\mu L_L^\beta) \right]~,
\label{eq:EFT}
\ee
where $v \approx 246$\,GeV. For simplicity, the definition of the EFT cutoff scale and the normalisation of the two operators is reabsorbed in the flavour-blind adimensional coefficients $C_S$~and~$C_T$.

The flavour structure in Eq.~(\ref{eq:EFT}) is contained in the Hermitian matrices $\llq_{ij}$, $\lle_{\alpha\beta}$ and follows from the assumed $U(2)_q\times U(2)_\ell$ flavour symmetry and its breaking. The flavour symmetry is defined as follows: the first two generations of left-handed quarks and leptons transform as doublets under the corresponding $U(2)$ groups, while the third generation and all 
the right-handed fermions are singlets.
Motivated by the observed pattern of the quark Yukawa couplings
(both mass eigenvalues and mixing matrix), it is further  assumed that the leading breaking terms of this flavour symmetry are two spurion doublets, $V_q$ and $V_\ell$, that give rise to the mixing between the third generation and the other two~\cite{Barbieri:2011ci,Barbieri:2012uh}.
The normalisation of $V_q$ is conventionally chosen to be $V_q \equiv  (V_{td}^*,V_{ts}^* )$, 
where $V_{j i}$ denote the elements of the Cabibbo-Kobayashi-Maskawa (CKM) matrix.
In the lepton sector we assume $V_\ell \equiv (0, V_{\tau\mu}^*  )$ with $|V_{\tau\mu} | \ll 1$.
We adopt as reference flavour basis the down-type quark  and  charged-lepton mass eigenstate basis,
where the $SU(2)_L$ structure of the left-handed fields is
\be
Q_L^i = \begin{pmatrix}V_{j i}^* u_L^j\\ d^i_L\end{pmatrix}~, \qquad 
L_L^\alpha = \begin{pmatrix} \nu_L^\alpha\\ \ell^\alpha_L\end{pmatrix}~.
\label{eq:basis0}
\ee

A detailed discussion about the most general flavour structure of the semi-leptonic operators 
compatible with the $U(2)_q\times U(2)_\ell$ flavour symmetry and the assumed symmetry-breaking 
terms is presented in Appendix~\ref{app:flavour}. The main points can be summarised as follows:
\begin{itemize}
\item[1.] The factorised flavour structure in  Eq.~(\ref{eq:EFT})
is not the most general one; however, it is general enough given that the available data are sensitive 
only to the flavour-breaking couplings $\llq_{sb}$ and $\lle_{\mu\mu}$ (and, to a minor extent, 
also to  $\lle_{\tau\mu}$). By construction, $\llq_{bb} = \lle_{\tau\tau} = 1$.
\item[2.]
The choice of basis in Eq.~(\ref{eq:basis0}) to define the $U(2)_q\times U(2)_\ell$ 
singlets (i.e.~to define the ``third generation" dominantly coupled to NP) is  arbitrary. 
This ambiguity reflects itself in the values of $\llq_{sb}$, $ \lle_{\mu\mu}$, and $\lle_{\tau\mu}$, 
 that, in absence of a specific basis alignment,  are expected to be 
\be
 \llq_{sb} = \mathcal{O}( |V_{cb}|)~, \quad 
 \lle_{\tau\mu}  =\mathcal{O}(|V_{\tau\mu}|)~, \quad
 \lle_{\mu\mu} =   \mathcal{O}(|V_{\tau\mu}|^2)~. 
\label{eq:U2est}
\ee
\item[3.]
A particularly restrictive scenario, that can be implemented both in the case of LQ or colour-less mediators,
is the so-called pure-mixing scenario, i.e.~the hypothesis that there exists a flavour basis where the NP interaction 
is completely aligned along the flavour singlets. For both mediators, in this specific limit one arrives to the prediction
$\lle_{\mu\mu}  > 0$.
\end{itemize}

In order to reduce the number of free parameters, in Eq.~(\ref{eq:EFT}) we assume the same flavour structure for the two operators. This condition is realised in specific simplified models, but it does not hold in general.  The consequences of relaxing this assumption are discussed in Section~\ref{sect:models} in the context of  specific examples. Finally, motivated by the absence of deviations from the SM in CP-violating observables, we assume all the complex phases, except the CKM phase contained in the $V_q$ spurion, to vanish (as shown in Appendix~\ref{app:flavour}, this implies $\llq_{bs}= \llq_{sb}$ and $\lle_{\tau\mu}= \lle_{\mu\tau}$).

\subsection{Fit of the semi-leptonic operators}
\label{sect:SLfit}

\begin{table}[t]
\begin{center}
\begin{tabular}{c|c|c}
	Observable & Experimental bound & Linearised expression \\ \hline
	\raisebox{-2pt}[8pt][8pt]{$\RD$} & \raisebox{-2pt}[8pt][8pt]{$1.237 \pm 0.053$} & \raisebox{-2pt}[8pt][8pt]{$1+2C_T (1 - \llq_{sb}  V_{tb}^* / V^*_{ts}  ) (1-\lle_{\mu\mu}/2)$} \\
	\raisebox{-2pt}[8pt][8pt]{$\Delta C_9^\mu=-\Delta C_{10}^\mu$} & $-0.61 \pm 0.12$ ~ \cite{Capdevila:2017bsm}& $- \frac{\pi}{\alpha_{\rm em} V_{tb} V_{ts}^*} \lle_{\mu\mu} \llq_{sb} (C_T+C_S) $ \\
	\raisebox{-2pt}[8pt][8pt]{$R_{b\rightarrow c}^{\mu e} - 1$} & \raisebox{-2pt}[8pt][8pt]{$0.00 \pm 0.02$} & $2 C_T (1 - \llq_{sb}  V_{tb}^* / V^*_{ts} ) \lle_{\mu\mu}$ \\
	\raisebox{-2pt}[8pt][8pt]{$B_{K^{(*)}\nu\bar\nu}$} & $0.0 \pm 2.6$ & $1+ \frac{2}{3}\frac{\pi}{\alpha_{\rm em} V_{tb} V_{ts}^* C_{\nu}^{\rm{SM}}} (C_T-C_S)  \llq_{sb} (1+\lle_{\mu\mu})$ \\
         \raisebox{-2pt}[8pt][8pt]{$\delta g_{\tau_L}^Z$} & $-0.0002 \pm 0.0006$ & $ 0.033 C_T - 0.043 C_S$ \\
         \raisebox{-2pt}[8pt][8pt]{$\delta g_{\nu_\tau}^Z$} & $-0.0040 \pm 0.0021$ & $ - 0.033 C_T - 0.043 C_S$ \\
	 \raisebox{-2pt}[8pt][8pt]{$|g^W_\tau / g^W_\ell|$} & $1.00097 \pm 0.00098$ & $1 - 0.084 C_T$ \\
	\raisebox{-2pt}[8pt][8pt]{$\mathcal{B}(\tau \to 3 \mu)$} & $ (0.0 \pm 0.6) \times 10^{-8} $ & $ 2.5 \times 10^{-4} (C_S - C_T)^2 (\lle_{\tau\mu})^2 $
\end{tabular}
\caption{\label{tab:FlavourFit} Observables entering in the fit, together with the associated experimental bounds (assuming the uncertainties follow the Gaussian distribution) and their linearised expressions in terms of the 
EFT parameters.  The full expressions used in the fit can be found in Appendix~\ref{app:obs}.}
\end{center}
\end{table}

To quantify how well the proposed framework can accommodate the observed anomalies, we perform a fit to low-energy data with four free parameters: $C_T$, $C_S$, $\llq_{sb}$, and $\lle_{\mu\mu}$, while for simplicity we set $\lle_{\tau\mu} = 0$.\footnote{We explicitly verified that a nonzero $\lambda_{\tau\mu}$ has no impact on the fit results.}
The set of experimental measurements entering the fit, together with their functional dependence on the fit parameters, is discussed in length in Appendix~\ref{app:obs}.
In particular, we take into account the LFU tests in the charged-current semi-leptonic observables $\RD$ and $R_{b\rightarrow c}^{\mu e}$, global fits of $b \to s\mu\mu$ processes (including the LFU ratios $\RK$ and  
the angular observables) along the direction $\Delta C_9^\mu = - \Delta C_{10}^\mu$ 
\cite{Altmannshofer:2017yso,DAmico:2017mtc,Capdevila:2017bsm,Ciuchini:2017mik,Geng:2017svp, Celis:2017doq, Hurth:2017hxg},
and limits on $\mathcal{B}(B \to K^* \nu\bar\nu)$ \cite{Buras:2014fpa}. We also include a set of observables sensitive to 
the purely-leptonic and electroweak operators generated by the renormalisation-group running of the semi-leptonic operators from the scale $\Lambda$ down to the electroweak scale.
The most notable effects are the corrections to the $Z\to \tau \bar\tau$ effective couplings, to the invisible $Z$ decay width, 
and to the LFU ($R_\tau^{\tau\ell}$) and LFV ($\tau \to 3 \mu$) tests in $\tau$ decays ~\cite{Feruglio:2016gvd,Feruglio:2017rjo}. 
The matching scale is set to $\Lambda=2$~TeV in the fit. The results change only slightly using $\Lambda=1$~TeV instead, 
relaxing the impact of the loop-induced constraints.
The observables considered in the fit 
are summarised in Table~\ref{tab:FlavourFit}, together with their approximate dependence on the EFT parameters.  
In order to fulfil the condition in Eq.~\eqref{eq:U2est} we impose $|\llq_{sb}| < 5 |V_{cb}| $.

\begin{figure}[tbp]
\centering%
\includegraphics[width=0.48\textwidth]{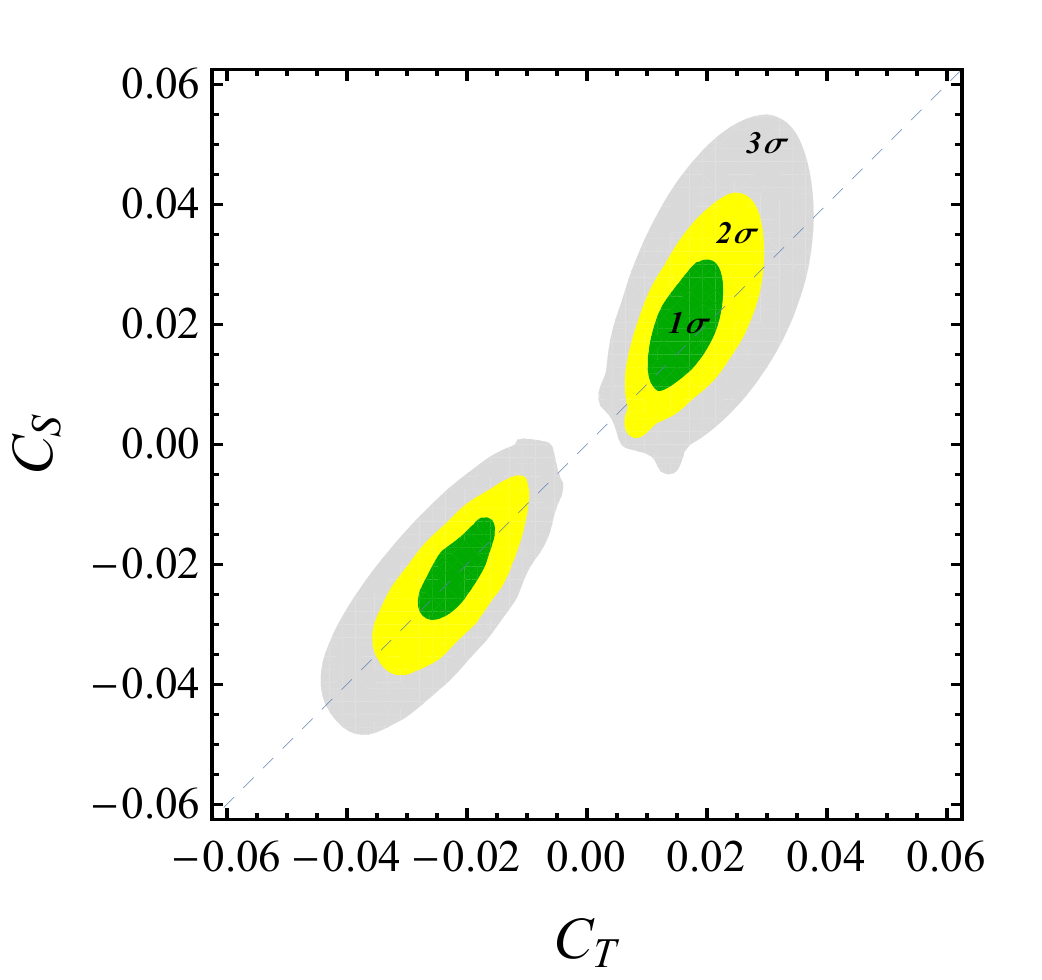} \hspace{0.5cm}
\includegraphics[width=0.455\textwidth]{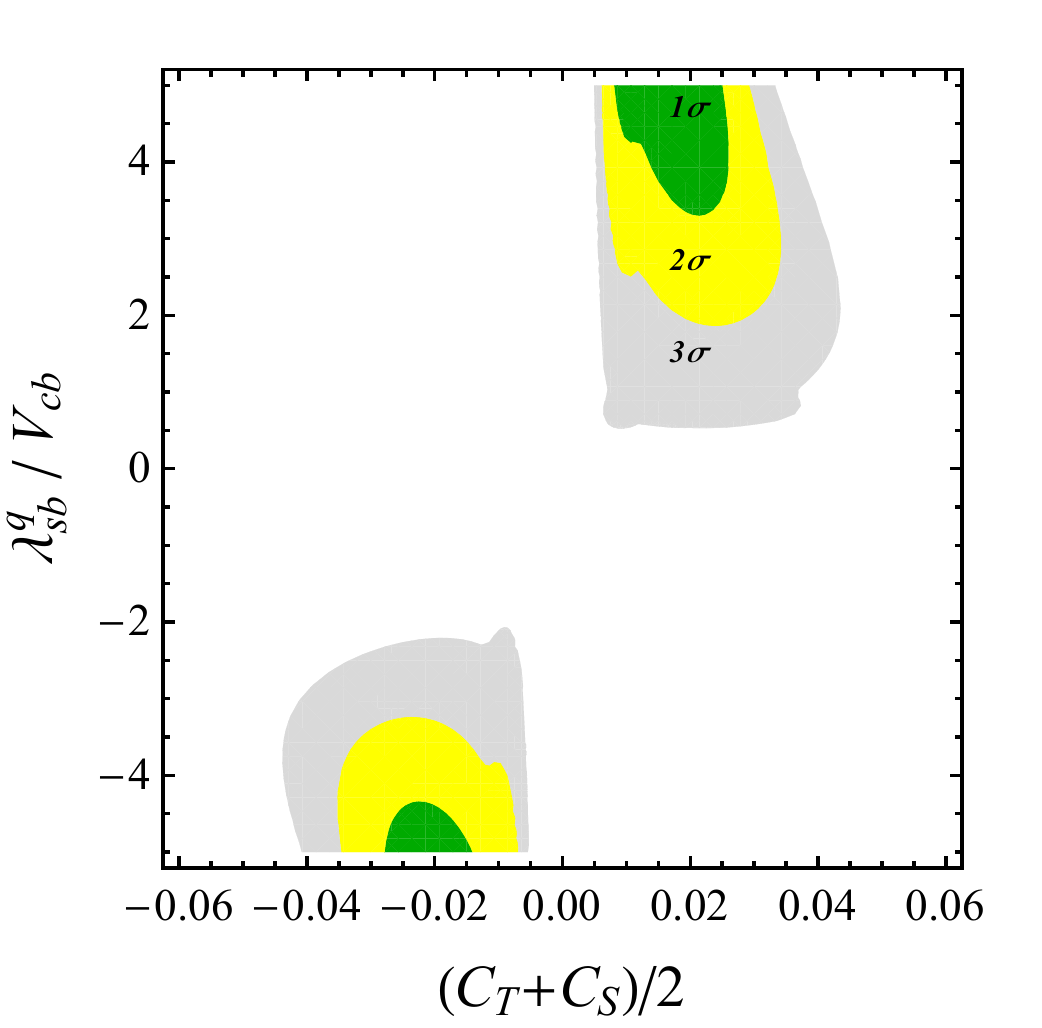}

\includegraphics[width=0.47\textwidth]{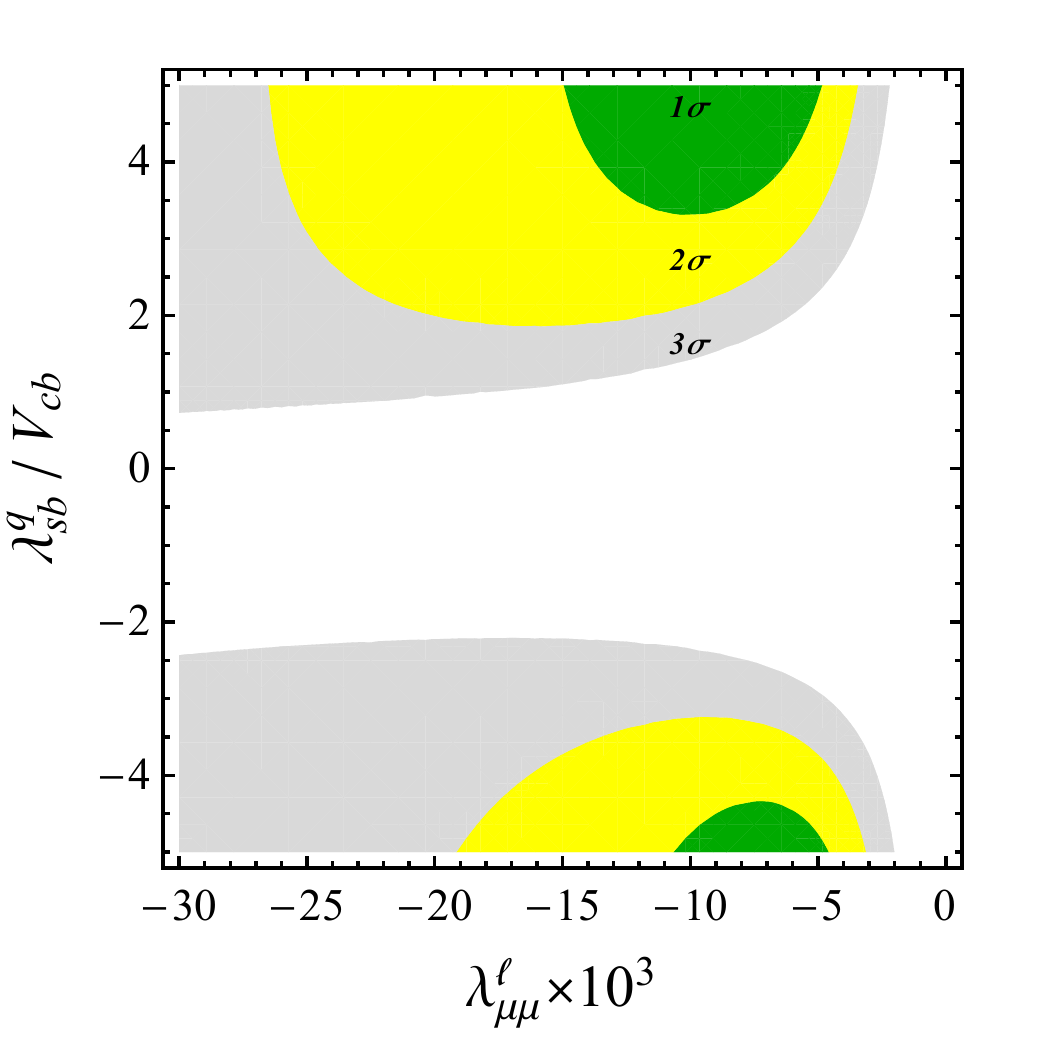} \hspace{0.5cm}
\includegraphics[width=0.47\textwidth]{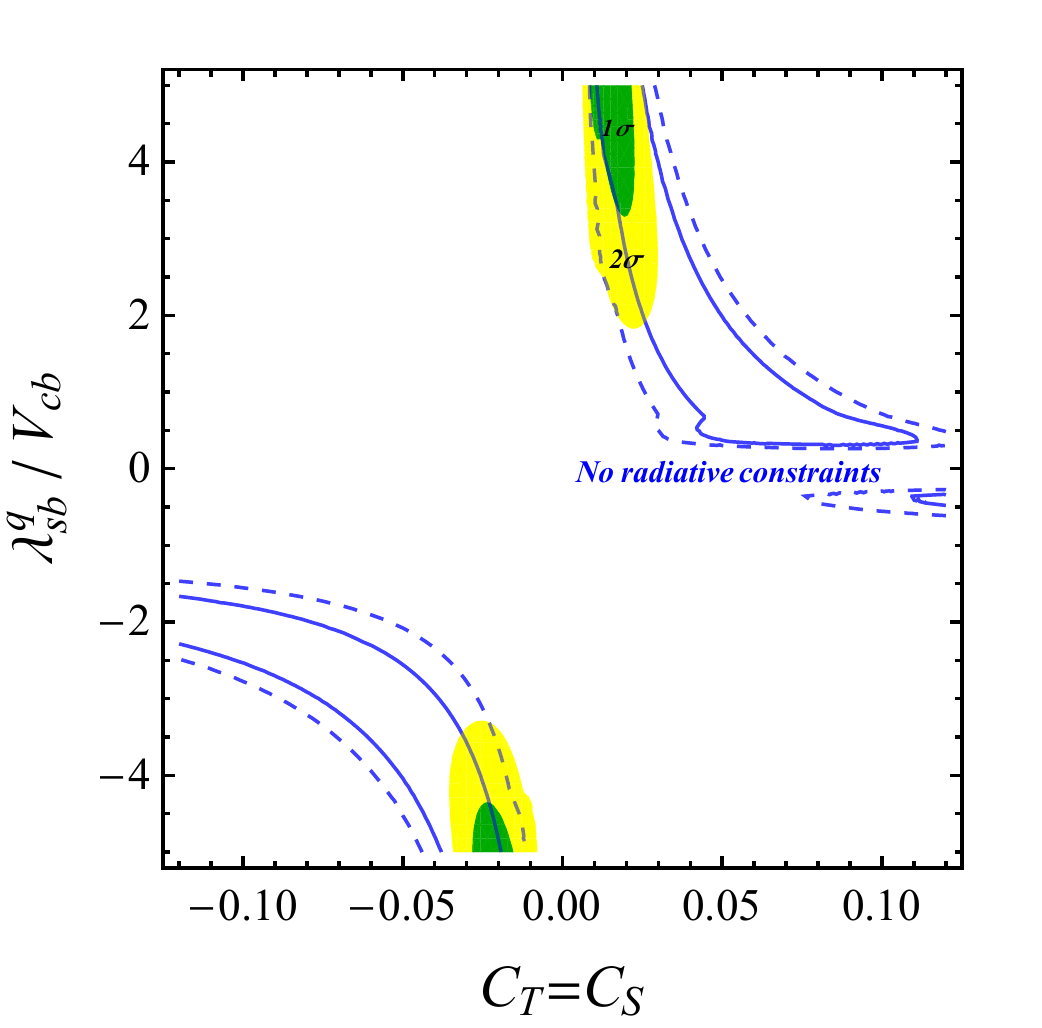}
\caption{\small Fit to the semi-leptonic and purely leptonic (radiatively generated) observables in Table~\ref{tab:FlavourFit}, in the framework of the triplet and singlet $V-A$ operators (see Eq.~\eqref{eq:EFT}), imposing $|\llq_{sb}| < 5 |V_{cb}|$.
In green, yellow, and gray, we show the $\Delta \chi^2 \leq$ 2.3 ($1\sigma$), 6.2 ($2\sigma$), and 11.8 ($3\sigma$) regions, respectively, after marginalising over all other parameters. In the bottom-right plot we fix $C_T = C_S$ and perform a fit with and without the radiatively induced observables.\label{Fig:fitEFT}}
\end{figure}

We minimise the total $\chi^2$ function to find the best-fit point and the corresponding confidence level intervals. The result are presented as 2D plots after marginalising over the other two parameters (see  Figure~\ref{Fig:fitEFT}). The main observations can be summarised as follows. 
\begin{itemize}
\item[1.] Because of radiative constraints, the fit favours sizeable values of $\llq_{sb}/V_{ts}^* \approx -\llq_{sb}/V_{cb}$, which allow to lower the value of  $C_{T,S}$ (i.e.~to increase the scale of NP) keeping fixed the contribution to $\RD$ (see the bottom-right panel of Figure~\ref{Fig:fitEFT}). This can be understood from the approximated expression for $\RD$ (see Appendix~\ref{app:obs} for the exact formula used in the numerical fit),
\be
	\RD \approx 1 + 2C_T \left(1 -  \llq_{sb} \frac{V_{tb}^* }{V^*_{ts} }\right)  = 1.237 \pm 0.053~,
\ee
where a smaller value for $C_T$ can be compensated by a larger one for $\llq_{sb}$.
The preferred values of $\llq_{sb}$ are still consistent with the general expectation in Eq.~\eqref{eq:U2est}. As we discuss below, the substantial increase in the effective NP scale is also beneficial in improving the agreement with the high-$p_T$ searches pointed out in~\cite{Faroughy:2016osc}.
\item[2.] The upper bound on $\mathcal{B}(B \to K^* \nu\bar\nu)$, as well as radiative constraints, strongly favour equal magnitudes of triplet and singlet operators ($C_T\sim C_S$).  Nevertheless, at the $1\sigma$ level this relation has to be satisfied only at the $30\%$ level,  and therefore requires no fine tuning.
\item[3.] The flavour symmetry plays a non-trivial role in avoiding significant constraints on the 
value of $\llq_{sb}$  from  $b\to u$ transitions, in particular  from $\cB(B\to \tau \nu)$, 
enforcing the relation $R^{\tau\ell}_{b\to u} = \RD$ (see Appendix~\ref{app:obs}).
 \item[4.]  The measured value of $\Delta C_9^\mu = - \Delta C_{10}^\mu$, together with the size of $\llq_{sb}$ and $C_{T,S}$ from points 1 and 2, requires a value of $\lle_{\mu\mu} \approx \mathcal{O}(10^{-2})$,  
 perfectly consistent with the hypothesis of a small breaking of the $U(2)_\ell$ flavour symmetry. 
 The measured values of  $\RK$   fix also the relative sign of $\lle_{\mu\mu}$ and $\lle_{\tau\tau}$
 which must be opposite,  strongly disfavouring the pure mixing hypothesis. 
\item[5.]
We do not include $\lambda_{\tau\mu}^\ell$ in the fit, but we point out that values of $|\lle_{\tau\mu}| \sim |\lle_{\mu\mu}|^{1/2} \sim 0.1$ are perfectly compatible with the limits from LFV in $\tau$ decays, even after taking into account radiatively-induced effects~\cite{Feruglio:2017rjo}.
We nevertheless list the related observable in Table~\ref{tab:FlavourFit} since it is relevant for some of the simplified models, 
 such as the scalar leptoquark, where $\lambda_{\tau\mu}^\ell$ cannot be set to zero.
\end{itemize}

\begin{figure}[tbp]
\centering
\includegraphics[width=0.48\textwidth]{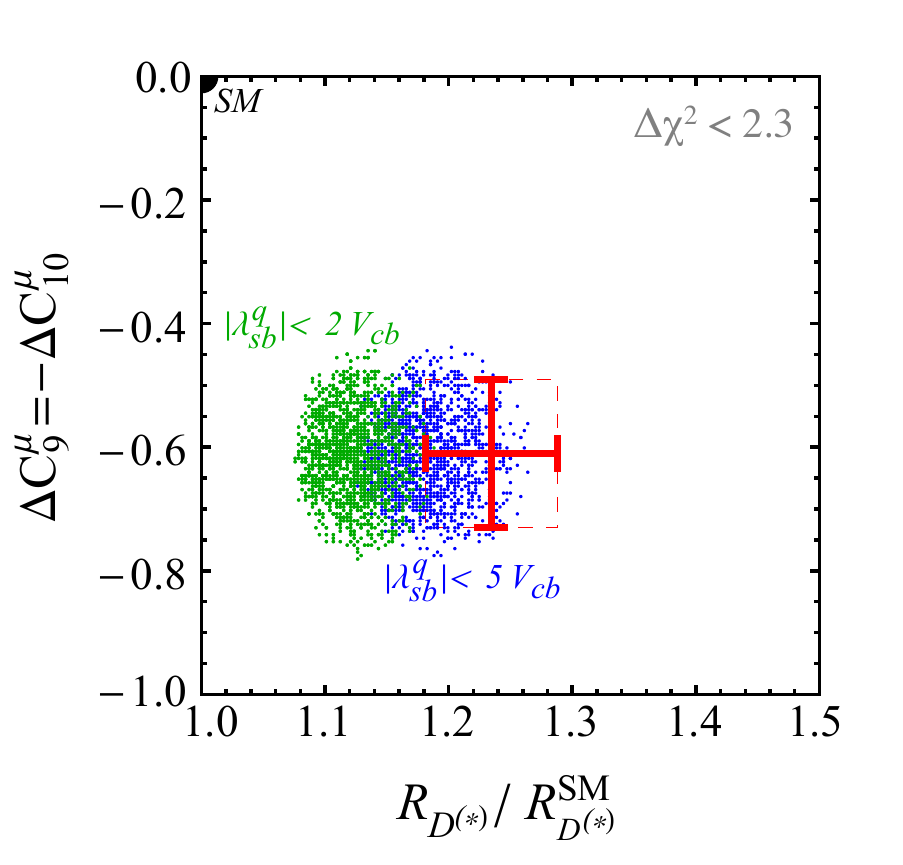}\hfill%
\raisebox{1.35em}{\includegraphics[width=0.405\textwidth]{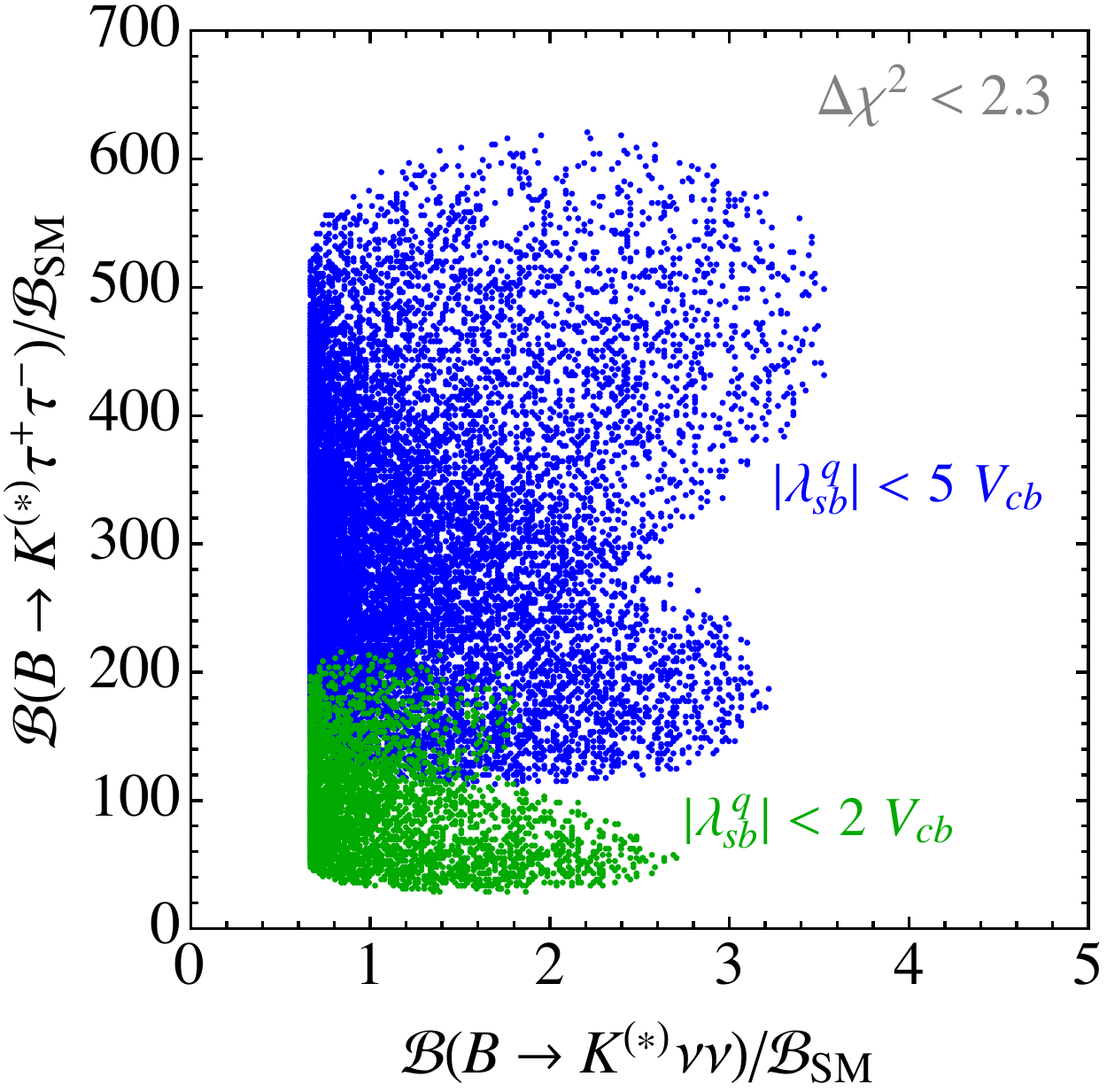}}
\caption{ \small  Left: Prediction for $\Delta C_9^\mu= - \Delta C_{10}^\mu$ (following from $\RK$) and $\RD$ for a randomly chosen set of points within the $1\sigma$ preferred region of the EFT fit: the blue points are obtained setting $|\llq_{sb}| < 5 |V_{cb}| $, while the green points are obtained setting the tighter condition $|\llq_{sb}| < 2 |V_{cb}|$ in the fit. The red cross denotes the 1$\sigma$ experimental constraint.
Right: expectations for $\cB(B \to K^{(*)}\nu\bar\nu)$ and  $\cB(B\to K^{(*)}\tau\bar\tau)$ within the 1$\sigma$ preferred values of the EFT fit, again for $\lambda_{sb}^q<5V_{cb}$ (blue) and $\lambda_{sb}^q<2V_{cb}$ (green).\label{Fig:fitEFTObs}}
\end{figure}

The best-fit region is consistent with both $\RK$ and $\RD$ anomalies.
To illustrate this fact, in Figure~\ref{Fig:fitEFTObs} we show the values of the two observables for a randomly chosen set of points within the $1\sigma$ preferred region ($\Delta \chi^2 < 2.3$). As can be seen, the upper bound set on $|\llq_{sb}|$ is strongly correlated to the maximal allowed NP contribution to $\RD$.

Analysing the correlations among the observables entering the fit, we find that more precise tests of 
LFU in $\tau$ decays and tighter constraints on the invisible $Z$ decay width would help 
in determining the sign of $C_T+C_S$.  
We also find a non-trivial correlation among the $Z\tau\bar\tau$ couplings and the $B\to K^{(*)}\nu\bar\nu$ branching ratio. 
These results motivate further tests of LFU in $Z$ and $\tau$ decays, as well the search for 
$b \to s \nu \bar \nu$ transitions. However, the smoking gun of the preferred solution of the EFT fit, 
that we denote the {\em large} $\llq_{bs}$ scenario,  is a huge enhancement of $b \to s \tau \bar \tau$ transitions
 -- between two and three orders of magnitude with respect to the SM -- as shown in Figure~\ref{Fig:fitEFTObs} (right). Such large values might be within the  experimental sensitivity of Belle II, which is expected to be of the order of $10^{-4}$ on the branching ratio \cite{Zupanc:2017CERN}. The size of the enhancement is clearly correlated with the maximal allowed value of $\lambda_{bs}$. 
The expected deviations from the SM in $R_{b\rightarrow c}^{\mu e}$ 
turn out to be well below the present sensitivity. 

\boldmath
\subsection{Beyond semi-leptonic operators: high-$p_T$ searches  and $\Delta F=2$}
\label{sect:beyondEFT}
\unboldmath

As we have shown, for reasonable values of the free parameters the effective Lagrangian in Eq.~(\ref{eq:EFT}) provides a good fit of  
both the $\RD$ and $b \to s \mu\mu$ anomalies,  being at the same time consistent with all available low-energy constraints.
The remaining two questions to address, which go beyond the simple EFT approach so far 
adopted, are the compatibility of the underlying model with high-$p_T$ searches, and bounds on pure-quark and pure-leptonic four-fermion operators.
Before analysing these questions in specific simplified models, it is worth trying to address 
them in general terms. 

As far as high-$p_T$ searches are concerned, particularly stringent bounds are set by $pp \to\tau\bar\tau + X$~\cite{Faroughy:2016osc}. While the form of the NP signal depends on the specific mediator (e.g.~colour-less vector or leptoquark), the overall strength is controlled by the values of $C_T$ and $C_S$ via the following effective interaction:
\be
\Delta \cL_{bb\tau\tau}  =  - \frac{1}{\Lambda^2_0}~ \left(\bar b_L  \gamma_\mu  b_L \right)   \left( \bar \tau_L \gamma_\mu   \tau_L \right)~, \qquad 
\qquad  \Lambda^2_0 =  \frac{v^2}{ C_S +C_T }~.
 \ee
The present bounds on the EFT scale $\Lambda_0$ were derived in \cite{Faroughy:2016osc} recasting different ATLAS searches for $\tau\bar\tau$ resonances, and read $\Lambda_0 > 0.62\,{\rm TeV}$. 
The fit discussed above implies $\Lambda_0 \approx 1.2$~TeV, which is well within the experimental limit.
Despite being a relatively low NP scale, 
this value is also high enough to pass the present constraints in most explicit models~\cite{Faroughy:2016osc}.\footnote{For comparison, 
the constraints derived in~\cite{Faroughy:2016osc} correspond to $C_T \approx 0.12$, which 
is about 6 (3) times larger than the best fit values of $C_T$ ($C_S+C_T$) 
in Figure~\ref{Fig:fitEFT}.} 
The EFT argument outlined above can only be taken qualitatively, as the validity of EFT expansion is expected to break at these energies.  Therefore,
a more detailed discussion on LHC limits is presented in the context of an explicit vector leptoquark model in Section~\ref{sec:vlq}. 
Another constraint on the size of $C_{S,T}$ comes from the study of perturbative unitarity in $2 \to 2$ scattering processes \cite{DiLuzio:2017chi}. Similarly to the one from direct searches, this bound is relevant for small $\llq_{bs}$ and large $C_{S,T}$, while it is easily satisfied in the region chosen by our EFT fit.

As far as other low-energy observables are concerned, the most problematic constraint is the one following 
from meson-antimeson mixing.
On  the one hand, given the symmetry and symmetry-breaking structure of the theory, 
we expect the underlying model to generate an effective 
interaction of the type 
\be
\Delta \cL_{(\Delta B = 2)}  =   C^{\rm NP}_0 \frac{(V_{tb}^*V_{ti})^2 }{32\pi^2 v^2} \left(\bar b_L  \gamma_\mu  d^i_L \right)^2~, \qquad  C^{\rm NP}_0 = \mathcal{O}(1) \times  \frac{32\pi^2 v^2}{\Lambda^2_0} 
\left| \frac{\llq_{sb}}{V_{cb}} \right|^2~.
\ee
The preferred values of  $\Lambda_0$ and $\llq_{sb}$
 from the EFT fit yield $C^{\rm NP}_0 = \mathcal{O}(100)$, while 
the experimental constraints
on $\Delta M_{B_{s,d}} $  require   $C^{\rm NP}_0$ to be at most 
$\mathcal{O}(10\%)$.
This problem poses a serious challenge to all models where 
$\Delta F=2$ effective operators are generated without some additional 
dynamical suppression compared to the semi-leptonic ones. 
A notable case where such suppression does occur are models with LQ mediators,
where $\Delta F=2$ amplitudes are generated only beyond the tree level.

An alternative to avoid the problem posed by $\Delta F=2$ constraints is to abandon 
the {\em large} $\llq_{sb}$ scenario preferred by the EFT fit, and assume  $|\llq_{sb}| \lesssim 0.1 \times |V_{cb}|$. 
In this limit the contribution to (down-type) $\Delta F=2$ amplitudes 
is suppressed also in presence of tree-level amplitudes. However, in order to cure the problem of the EFT fit, in this 
case one needs additional contributions to compensate for the radiative constraints  (see Figure~\ref{Fig:fitEFT}   bottom-right).  
In other words, in the  {\em small} $\llq_{sb}$ scenario the tuning problem is moved from 
the $\Delta F=2$ sector to that of electroweak observables. We will present an explicit realisation of the small $\llq_{sb}$ scenario in Section~\ref{sect:vectors}.

\begin{figure}[tbp]
\centering%
\includegraphics[width=0.5\textwidth]{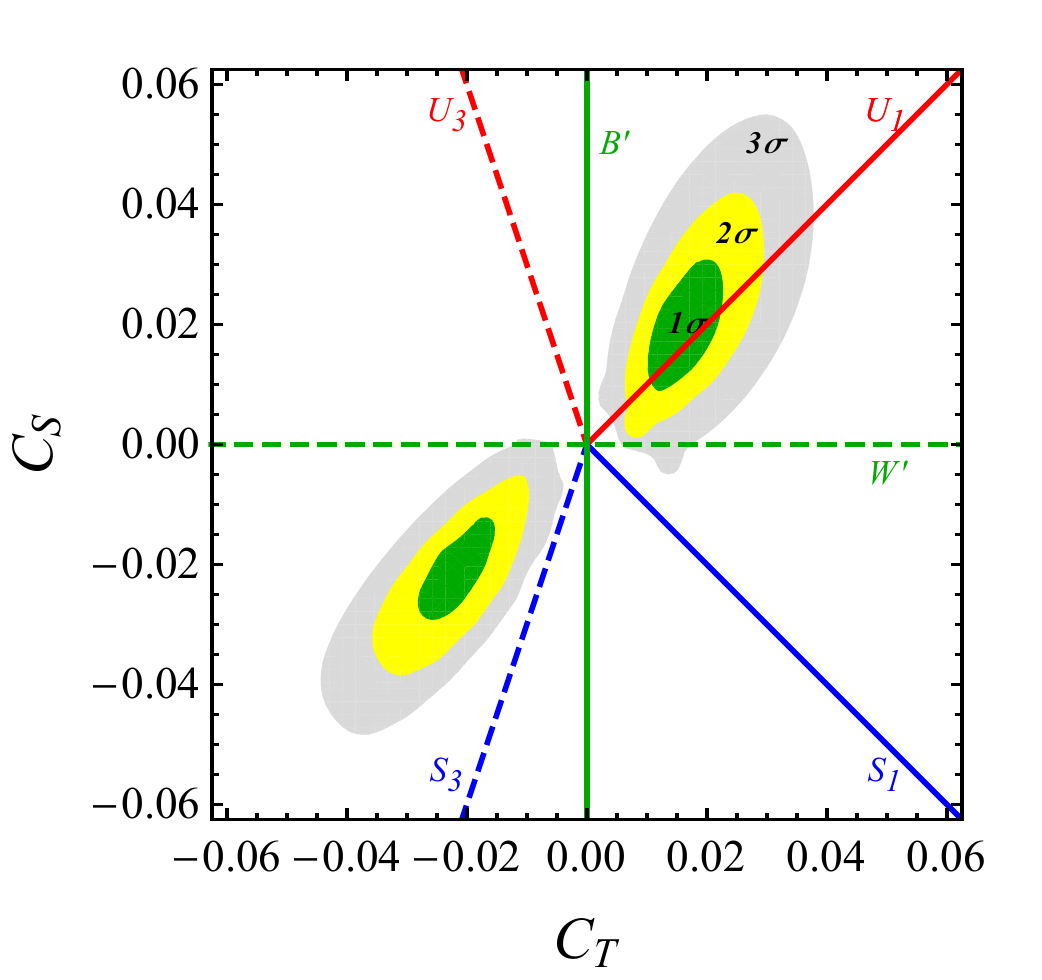}
\caption{ \small  The lines show the correlations among triplet and singlet operators in single-mediator models. Colour-less vectors are shown in green, coloured scalar in blue, while coloured vectors in red. Electroweak singlet mediators are shown with the solid lines while triplets with dashed.
\label{Fig:SingleRes} }
\end{figure}

\section{Simplified models}
\label{sect:models}

In this section we analyse how the general results discussed in the previous section can be implemented, 
and eventually modified adding extra ingredients, in three specific (simplified) UV scenarios with explicit mediators.

The complete set of single-mediator models with tree-level matching to the vector triplet and/or singlet $V-A$ operators
consists of: colour-singlet vectors $B'_\mu \sim ({\bf 1},{\bf 1},0)$ and $W'_\mu \sim ({\bf 1},{\bf 3},0)$,~colour-triplet scalars $S_1 \sim (\bar {\bf 3},{\bf 1},1/3)$ and  $S_3 \sim (\bar {\bf 3},{\bf 3},1/3)$, and~coloured vectors $U_1^\mu \sim ({\bf 3},{\bf 1},2/3)$ and $U_3^\mu \sim ({\bf 3},{\bf 3},2/3)$~\cite{Dorsner:2016wpm}. The quantum numbers in brackets indicate colour, weak, and hypercharge representations, respectively. 
In Figure~\ref{Fig:SingleRes} we show the correlation between triplet and singlet operators 
predicted in all single-mediator models, compared to the regions favoured by the EFT fit. 

The plot in Figure~\ref{Fig:SingleRes}  clearly singles out the case of a vector LQ, $U_1^\mu$,  which we closely examine in the next subsection, 
as the best single-mediator case. However, it must be stressed that there is no fundamental reason to expect the low-energy anomalies to be saturated by the contribution of a single tree-level mediator.
In fact, in many UV completions incorporating one of these mediators (for example in composite Higgs models, see Section~\ref{sec:UV}), 
these states often arise with partners of similar mass but different electroweak representation, and it is thus natural to consider two 
or more of them at the same time. For this reason, and also for illustrative purposes, in the following subsections we 
consider two representative cases with more than one mediator at work: two colour-less vectors, $SU(2)_L$ triplet and singlet, and two coloured scalars, also electroweak triplet and singlet. 

\begin{figure}[tbp]
\centering%
\includegraphics[width=0.45\textwidth]{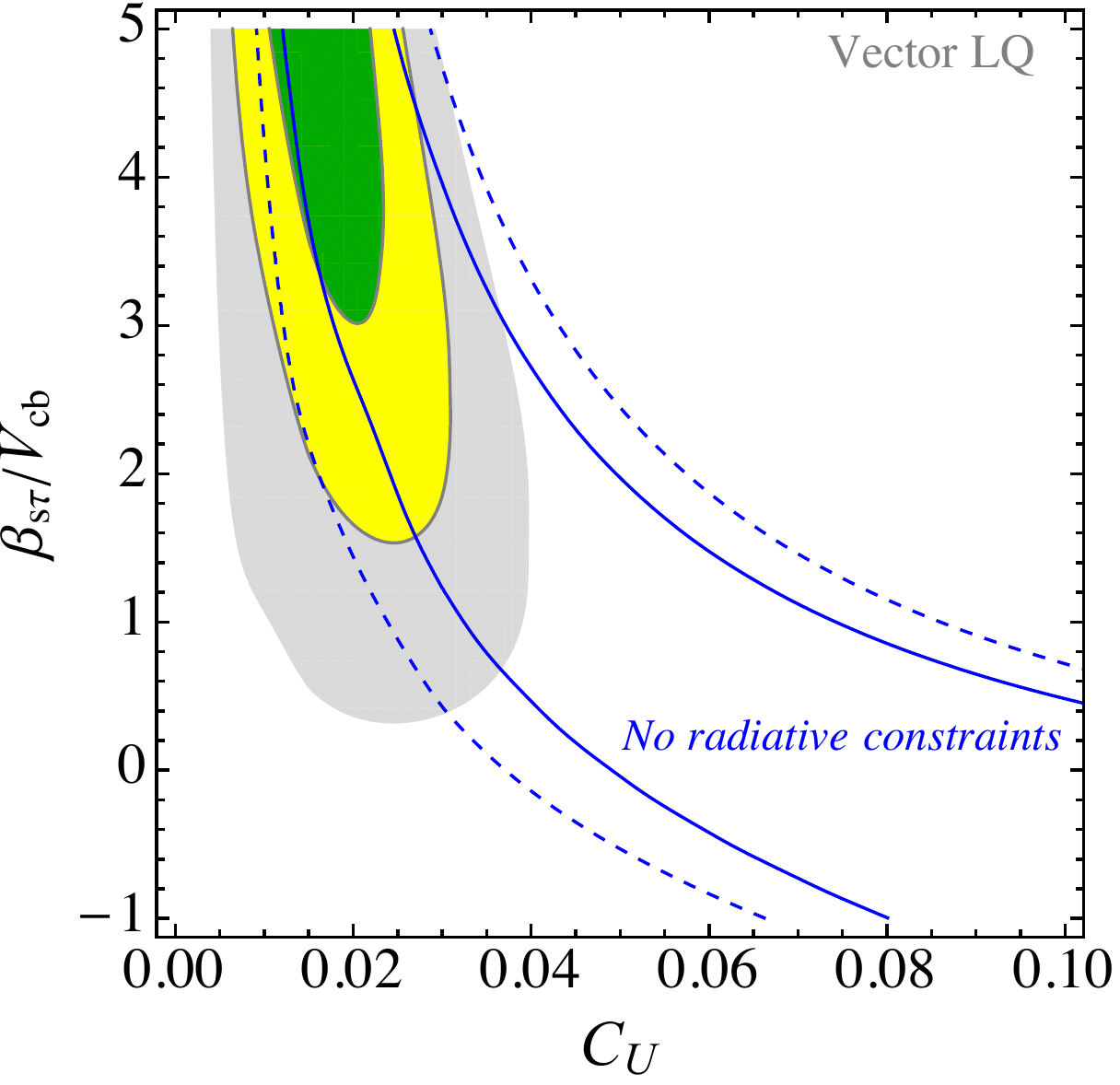} \quad
\includegraphics[width=0.46\textwidth]{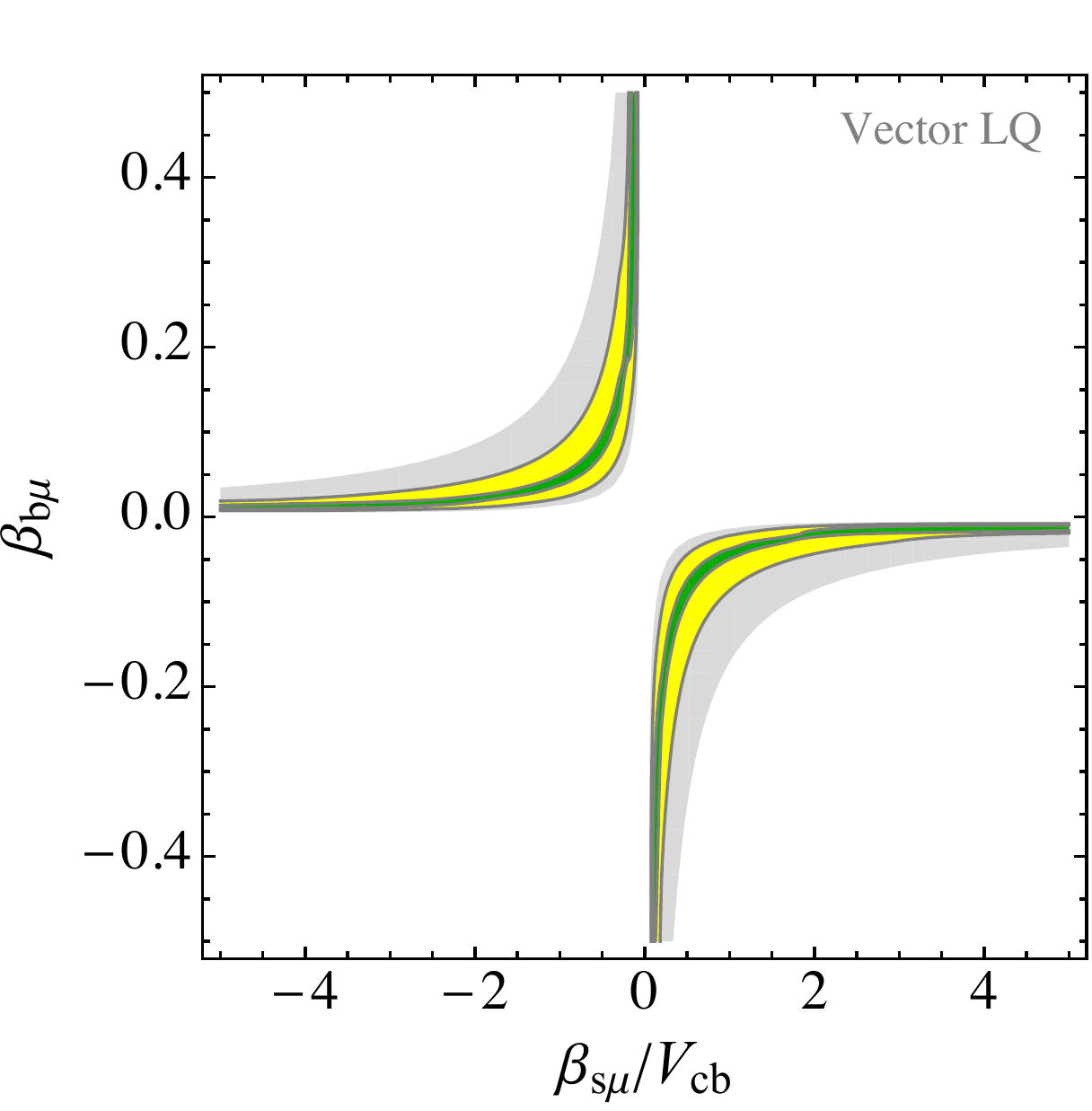}
\caption{ \small  Fit to semi-leptonic and radiatively-generated purely leptonic observables in Table~\ref{tab:FlavourFit}, for the vector leptoquark $U_\mu$, imposing $|\beta_{s\mu,s\tau}| < 5 |V_{cb}|$  and $C_U > 0$. 
In green, yellow, and gray, we show the $\Delta \chi^2 \leq$ 2.3 ($1\sigma$), 6.0 ($2\sigma$), and 11.6 ($3\sigma$) regions, respectively. The dashed and solid blue lines represent the $1$ and $2\sigma$ limits in the case where radiative constraints are removed from the fit. 
\label{Fig:LQfit} }
\end{figure}

\subsection{Scenario I: Vector Leptoquark}
\label{sec:vlq}

As anticipated, the simplest UV realisation of the scenario emerging from the EFT fit 
is that of an $SU(2)_L$-singlet vector leptoquark, $U^\mu_1 \equiv ({\bf 3},{\bf 1},2/3)$, coupled to the left-handed quark and lepton currents
\ba
\mathcal L_U &=&  ~-\frac{1}{2} U^\dagger_{1,\mu\nu} U^{1,\mu \nu} +M_U^2 U^\dagger_{1,\mu} U_1^\mu+ g_U (J^\mu_{U} U_{1,\mu}+\rm{h.c.})~,\\
J^\mu_{U} &\equiv&  \beta_{i \alpha}~ \bar Q_i \gamma^\mu L_\alpha\,~.
\ea
Here $\beta_{i \alpha}^{(0)} =  \delta_{3 i} \delta_{3 \alpha}$ up to
$U(2)_q \times U(2)_\ell$ breaking terms, as shown in Eq.~(\ref{eq:currents}), and the 
flavour structure used in the general fit is recovered by means of the relations  (\ref{eq:LQbeta}).
After integrating out the leptoquark field, the tree-level matching condition for the EFT is 
\be
	\mathcal{L}_{\rm eff} \supset - \frac{1}{v^2} ~C_U~ \beta_{i\alpha} \beta^*_{j\beta} \left[ (\bar Q_L^i \gamma_\mu \sigma^a Q_L^j ) (\bar L_L^\beta \gamma^\mu \sigma^a L_L^\alpha) + (\bar Q_L^i \gamma_\mu Q_L^j) (\bar L_L^\beta \gamma^\mu L_L^\alpha) \right]~,
\ee
where $C_U = v^2 |g_U|^2/ (2 M_U^2) > 0$. Note that in this case the singlet and triplet operators have the same flavour structure and, importantly, the relation 
$C_S=C_T$ is automatically fulfilled  at the tree-level.  Furthermore, as already stressed, 
the flavour-blind contraction involving light fermions (flavour doublets) is automatically forbidden by the $U(2)_q \times U(2)_\ell$ symmetry. Last but not least, this LQ representation does not allow baryon number violating operators of dimension four.
These features, and the absence of a tree-level contribution to $B_{s(d)}$  meson-antimeson 
mixing, makes this UV realisation, originally proposed in \cite{Barbieri:2015yvd}, particularly appealing: the best fit points of
the general fit in Section~\ref{sect:SLfit} can be recovered essentially without tuning of the model 
parameters.

In Figure~\ref{Fig:LQfit} we show the results of the flavour fit in this parametrisation (using the $\beta_{i\alpha}$ rather than the $\lambda^{q(\ell)}_{ij(\alpha\beta)}$ as free parameters).
When marginalising we let $\beta_{s\tau}$ and $\beta_{s\mu}$ vary between $\pm 5 |V_{cb}|$ and impose $|\beta_{b\mu}| < 0.5$. 
We find very similar conclusions to the previous fit, in particular a reduced value of $C_U$ thanks to the extra contribution to $\RD$ proportional to $\beta_{s\tau}$,   
with both this parameter and $\beta_{s\mu}$ of $\mathcal{O}(|V_{cb}|)$. 

\begin{figure}[tbp]
  \centering
      \includegraphics[width=0.6\textwidth]{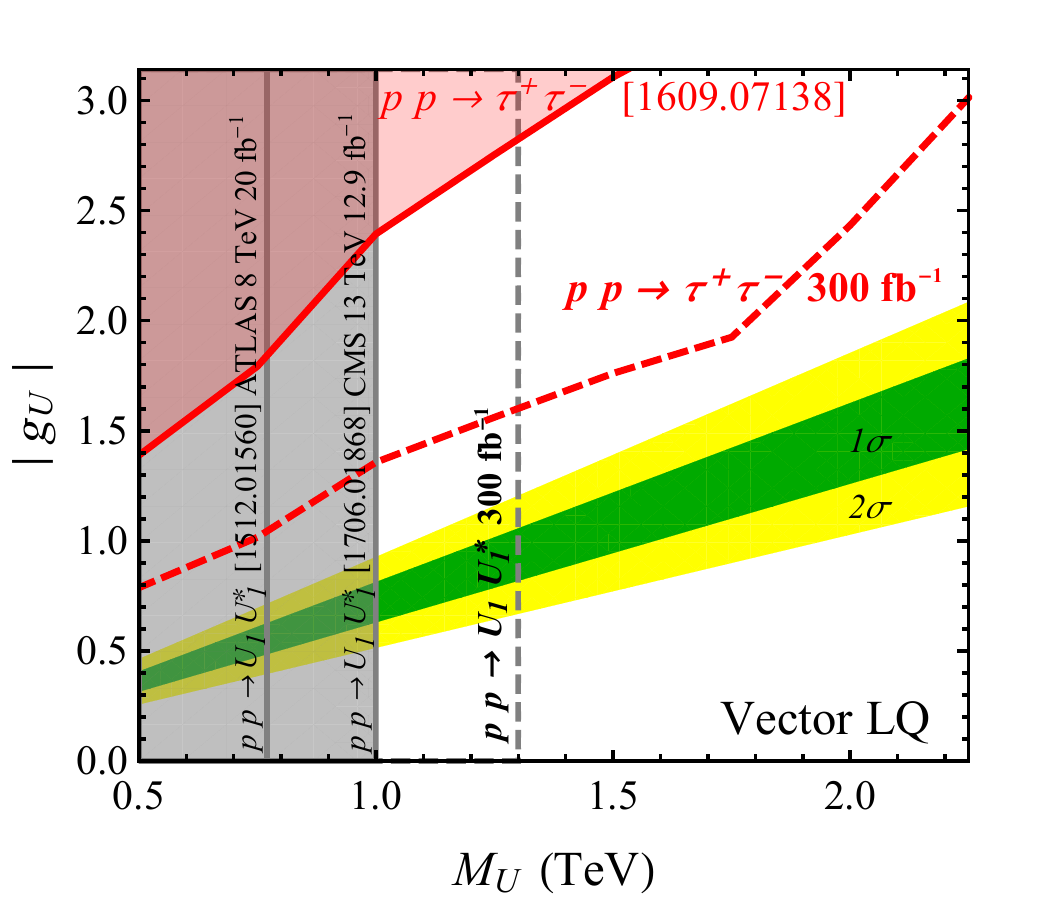}
    \caption{ \small  Present and future-projected LHC constraints on the vector leptoquark model of Section~\ref{sec:vlq}. The $1\sigma$ and $2\sigma$ preferred regions from the low-energy fit are shown in green and yellow, respectively.
    \label{Fig:ColliderU1} }
\end{figure}

Despite being absent at the tree level, a contribution to $\Delta F=2$ amplitudes is generated in this model 
at the one-loop level. The result thus obtained is quadratically divergent and  therefore strongly dependent 
on the UV completion. Following the analysis of Ref.~\cite{Barbieri:2015yvd}, 
i.e.~setting a hard cut-off $\Lambda$ 
on the quadratically divergent $\Delta F=2$ (down-type) amplitudes, leads to 
\be
\Delta \cL_{(\Delta B = 2)}   =  C^{(U)}_0 \frac{(V_{tb}^*V_{ti})^2 }{32 \pi^2 v^2}~ \left(\bar b_L  \gamma_\mu  d^i_L \right)^2~, \qquad  C^{(U)}_0 =  C^2_U \left(\frac{ \llq_{bs}}{V_{ts}}\right)^2 \frac{\Lambda^2}{ 2 v^2}~.
\label{eq:DF2U}
\ee
As already pointed out in Section~\ref{sect:beyondEFT}, the value of  $C^{(U)}_0$
should not exceed $\mathcal{O}(10\%)$ given the experimental constraints 
on $\Delta M_{B_{s,d}}$ (for comparison, $C^{(\rm SM)}_0 = (4\pi\alpha/s^2_W) S_0(x_t) \approx 1.0$,
see Appendix~\ref{app:obs}). This can be achieved only 
for $\Lambda \sim$ few TeV -- i.e.~$\Lambda$ not far from  $M_U$, as expected 
in a strongly interacting regime (unless some specific cancellation mechanism 
of $\Delta F=2$  amplitudes is present in the UV).
Interestingly enough, for fixed $\Lambda$, the large value of 
$\llq_{bs}$ does not increase the tension (contrary to the colour-less vector case discussed in
Section~\ref{sect:vectors}) due to the quadratic dependence on $C_U$ in Eq.~(\ref{eq:DF2U}).

\subsubsection*{High-energy constraints and strategies for direct searches}

Vector leptoquarks are copiously produced in pairs at the LHC due to QCD interactions. Unlike scalar LQ pair production, the theory prediction for the production cross-section in the vector case is less robust and depends on the size of non-minimal couplings to gluons. Nevertheless, the minimal coupling scenario gives rather conservative estimates of the production cross-section, which is roughly a factor of 10 larger than for the scalar LQ of the same mass~\cite{Belyaev:2005ew}.
Due to the flavour structure specified above, the $U_1^\mu$ leptoquark is expected to decay to $t \bar\nu_\tau$ and $b \bar\tau$ final states democratically. The CMS collaboration has searched for scalar LQ produced in pairs and decaying to these final states with $19.7$~fb$^{-1}$ at 8~TeV~\cite{Khachatryan:2014ura}. The results are reported in Figure~5 of~\cite{Khachatryan:2014ura}, showing the comparable sensitivity in the two channels for our scenario with $\mathcal{B}(U_1^\mu \to t \bar\nu_\tau)=\mathcal{B}(U_1^\mu \to b \bar\tau)=0.5$.
Similar limits in the $t \bar t \nu \bar\nu$ channel are reported by the ATLAS collaboration using the 8~TeV dataset~\cite{Aad:2015caa}. Assuming the same efficiencies and correcting for the production cross-section and branching ratio, the lower limit on the vector LQ mass is set to $M_U > 770$~GeV~\cite{Barbieri:2015yvd}. {Similarly, recent search by CMS at 13 TeV with 12.9~fb$^{-1}$~\cite{Sirunyan:2017yrk} implies $M_U > 1.0$~TeV~\cite{DiLuzio:2017chi}.} Naively rescaling these limits with the luminosity and cross-section at 13~TeV, the LHC reach with $300$~fb$^{-1}$ is about $1.3$~TeV. 

Another relevant collider signature is the production of tau lepton pairs at high energies ($pp \to\tau\bar\tau + X$)  due to the
$t-$channel (tree-level) leptoquark exchange. A recast of the ATLAS search~\cite{Aad:2015osa} already sets relevant bounds for the vector leptoquark explanation of the $R(D^{*})$ anomaly in the limit $\beta_{s\tau} \to 0$~\cite{Faroughy:2016osc} (in this limit, the radiative constraints in the $Z$ and lepton sector are to be addressed by some other mechanism, for example by a mild tuning with other contributions).
Instead, we find that with the value of $\beta_{s\tau} =$~(few)~$\times V_{cb}$, naturally emerging from the fit after the inclusion of radiative constraints, these 
bounds are easily satisfied. The preferred value of the fit requires $C_U$ to be about $7$ times smaller than what is obtained in the $\beta_{s\tau} \to 0$ limit, 
implying that the $b \bar b \to \tau \bar \tau$ production signal drops by almost a factor of $50$.  
The $s \bar b  (\bar s b) \to \tau \bar \tau$ and 
$s \bar s \to \tau \bar \tau$ production cross-sections are instead sub-leading since the enhancement due to the strange quark parton distribution 
function does not compensate for the $|\beta_{s\tau}|^2$ ($|\beta_{s\tau}|^4$) suppression.

The compilation of the leading collider bounds, as well as the corresponding projections for $300$~fb$^{-1}$, is shown in Figure~\ref{Fig:ColliderU1}. The preferred range of $C_U$ from the fit in Figure~\ref{Fig:LQfit} is translated to the green ($1\sigma$) and yellow ($2\sigma$) bands in Figure~\ref{Fig:ColliderU1}. This is a striking example of a scenario that 
could require HL-LHC (or significantly optimised search strategies) in order to obtain a high-$p_T$ signature of the mediator responsible to the $B$-physics anomalies.
 
As a final remark, collider signatures involving muons in the final state~\cite{Greljo:2017vvb} can be relevant in the future for large values of the $\beta_{b \mu}$ parameter corresponding to the corners of the preferred region shown in Figure~\ref{Fig:LQfit}. In this respect, LQ pair production decaying into the $b \tau b \mu$ final state (or even $b\mu b \mu$~\cite{Diaz:2017lit}), 
as well as single LQ  production in association with a muon~\cite{Dorsner:2014axa}, are potentially interesting search modes for 
this framework.

\subsection{Scenario II: Scalar Leptoquarks}
\label{sec:2LQ}

We introduce two scalar leptoquarks $S_1=(\overline{\mathbf{3}},\mathbf{1},1/3)$ and $S_3= (\overline{\mathbf{3}},\mathbf{3},1/3)$. The relevant interaction Lagrangian is given by~\cite{Dorsner:2016wpm}
\begin{align}
\label{eq:main_S_3}
\mathcal{L} \supset &~ g_1 {\beta_{1,i\alpha}} (\bar{Q}_{L}^{c\,i} \epsilon L_{L}^{\alpha}) S_{1} + g_3 
{\beta_{3,i \alpha}} (\bar{Q}_{L}^{c\,i} \epsilon \sigma^a L_{L}^{\alpha} ) S^a_{3}+\textrm{h.c.},
\end{align}
where $\epsilon=i \sigma^2$, $Q^c_L = C \bar{Q}_L^T$, and $S^a_{3}$ are the components of the $S_{3}$ leptoquark in $SU(2)_L$ space.
A model with the same field content was recently proposed in~\cite{Crivellin:2017zlb} as a possible solution of the $B$-physics anomalies.
However, the  flavour structure postulated in~\cite{Crivellin:2017zlb} leads to large cancellations 
in $b\to s \nu \bar\nu$ and potential tuning also in  $b\to u$ charged-current transitions.
Contrary to the vector LQ case, 
baryon number conservation is not automatically absent in the renormalisable 
operators built in terms of $S_{1,3}$
and must be imposed as an additional symmetry of the theory. 

\begin{figure}[tbp]
\centering%
\includegraphics[width=0.46\textwidth]{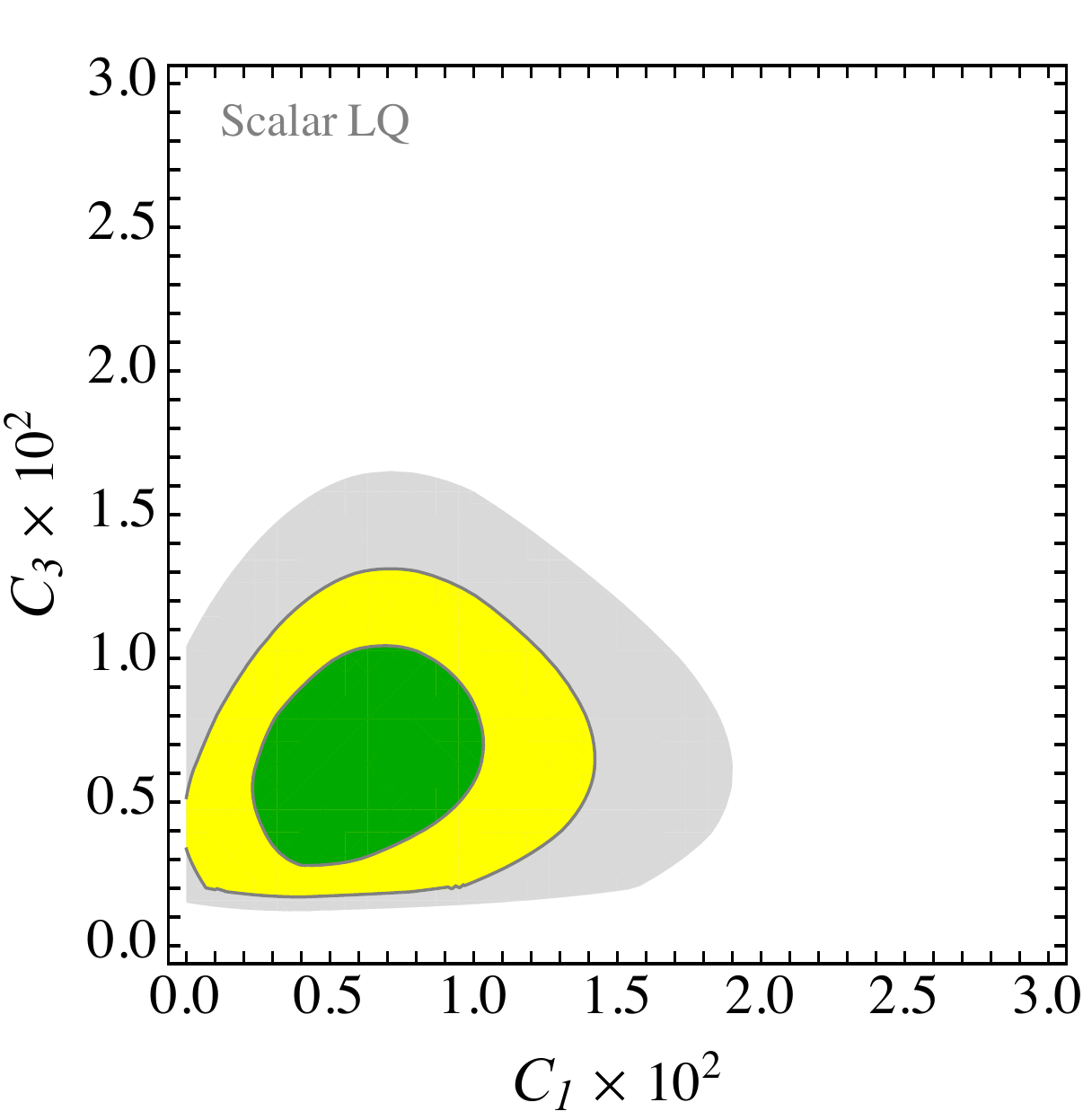}\hfill%
\includegraphics[width=0.45\textwidth]{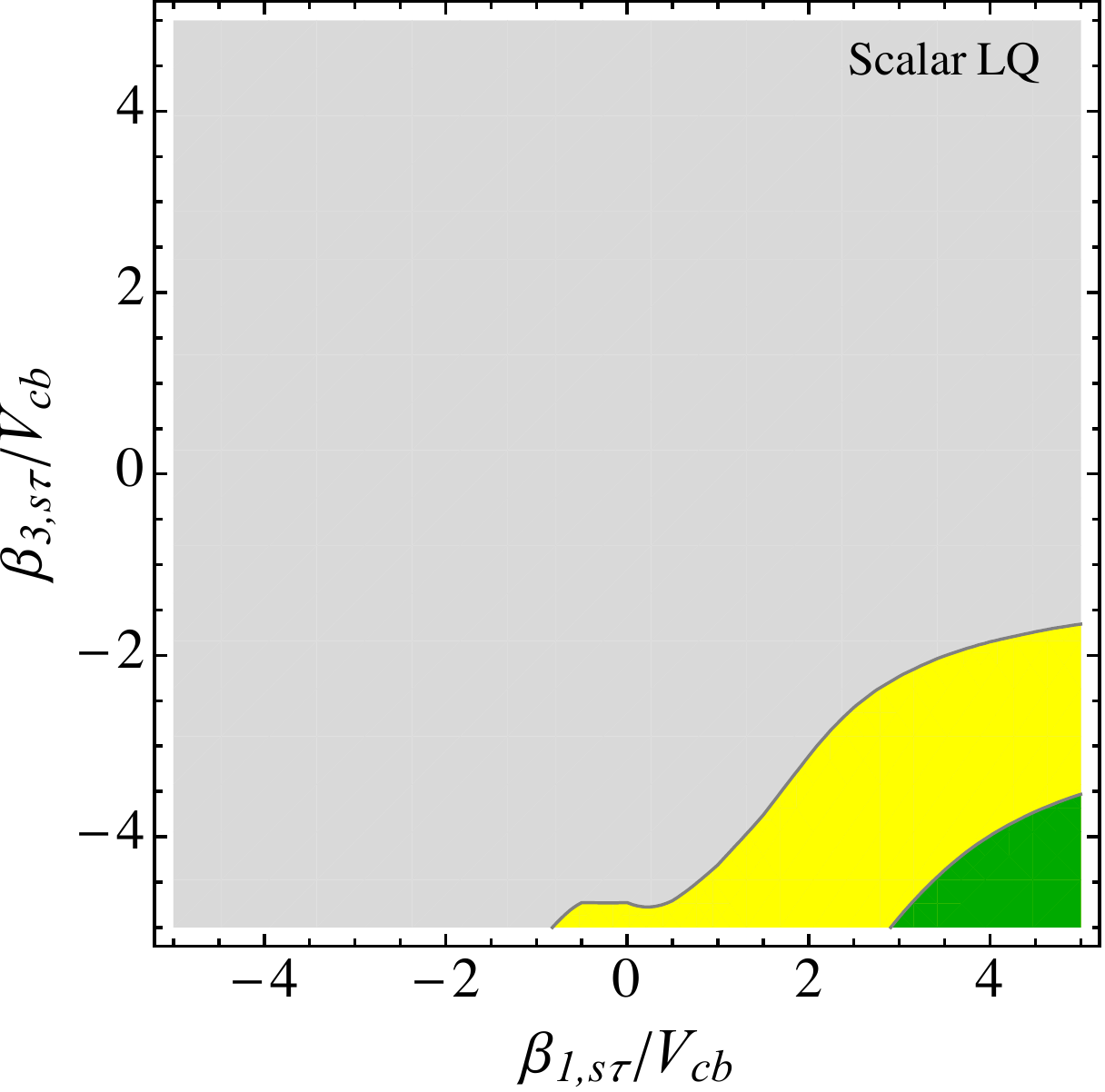}  

\vspace{0.5cm}

\includegraphics[width=0.45\textwidth]{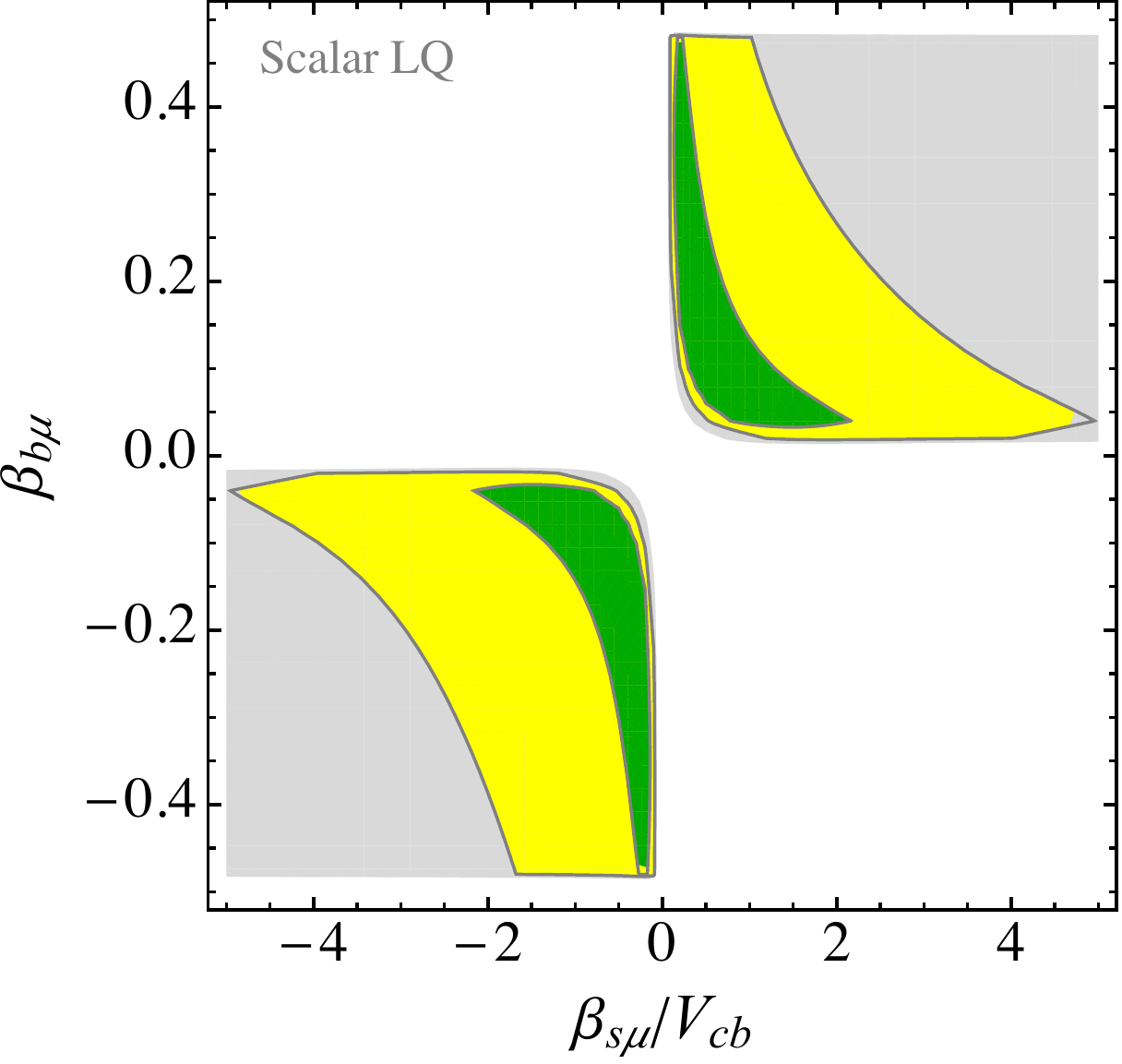}\hfill%
\includegraphics[width=0.45\textwidth]{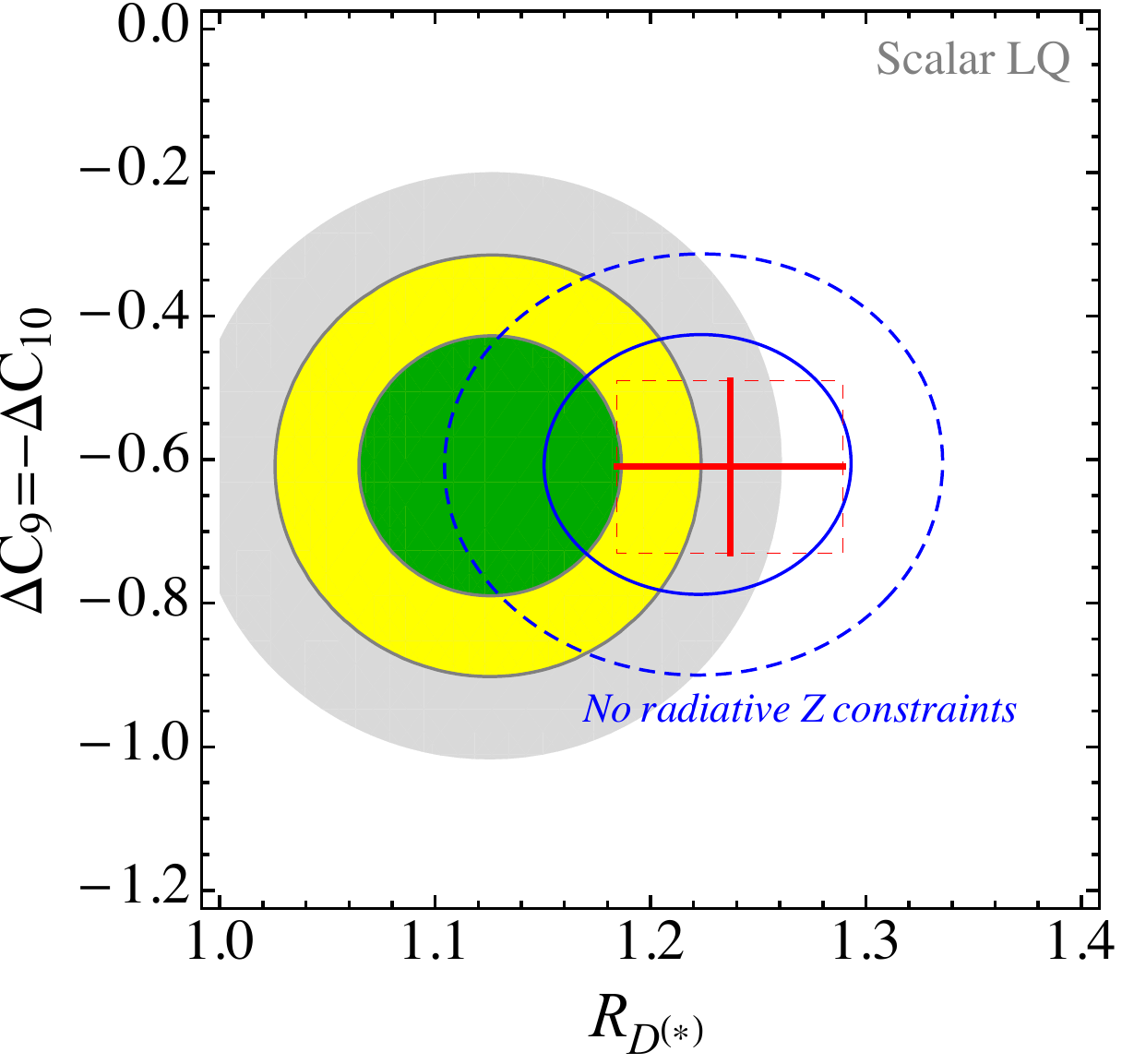}
\caption{ \small Fit to the semi-leptonic and radiatively-generated purely leptonic observables in Table~\ref{tab:FlavourFit}, for the scalar leptoquarks $S_1$ and $S_3$, imposing $|\beta_{s\mu,s\tau}| < 5 |V_{cb}|$ and $C_{1,3} > 0$. 
In green, yellow, and gray, we show the $\Delta \chi^2 \leq$ 2.3 ($1\sigma$), 6.2 ($2\sigma$), and 11.8 ($3\sigma$) regions, respectively.
In the lower-right panel we show the preferred values of the fit in the $R_{D^(*)}$, $\Delta C^\mu_9$ plane, compared with the $1\sigma$ experimental measurements (red box). Removing $Z\to\tau\bar\tau,\nu\bar\nu$ radiative constraints from the fit, the 1- and $2\sigma$ preferred regions in this case are shown with solid and dashed blue lines.
\label{Fig:SLQ} }
\end{figure}

Integrating out the leptoquark states at tree-level and matching to the effective theory, we find the following semi-leptonic operators
\begin{equation}
\begin{aligned}
 \mathcal{L}_{\rm{eff}} & \supset - \frac{1}{v^2}\left( C_1 \beta_{1, i \beta} \beta^{*}_{1, j \alpha} - C_3 \beta_{3, i \beta} \beta^{*}_{3, j \alpha}\right)~(\bar Q_L^i \gamma_\mu \sigma^a Q_L^j ) (\bar L_L^\alpha \gamma^\mu \sigma^a L_L^\beta) \\
  &- \frac{1}{v^2} \left(- C_1 \beta_{1, i \beta} \beta^{*}_{1, j \alpha} - 3 C_3 \beta_{3, i \beta} \beta^{*}_{3, j \alpha} \right)~ (\bar Q_L^i \gamma_\mu Q_L^j) (\bar L_L^\alpha \gamma^\mu L_L^\beta) ~,
 \end{aligned}
\end{equation}
where {$C_{1,3} = v^2 |g_{1,3}|^2/(4 M_{S_{1,3}}^2) > 0$.} Enforcing a minimally broken $U(2)_q \times U(2)_\ell$ flavour symmetry 
the two mixing matrices $\beta_{1,i\alpha}$ and $\beta_{3,i\alpha}$ follow the decomposition presented 
in Appendix~\ref{app:flavour} and have a hierarchical structure similar to the $\beta_{i\alpha}$ of the vector LQ case.  These two flavour matrices are, in general, different. However, for the sake of simplicity, 
in the fit we fix $\beta_{3,s\mu} = \beta_{1,s\mu}$ and $\beta_{1,b\mu} = \beta_{3,b\mu}$, keeping only the two $s-\tau$ elements different (since this is required for the fit to work).
The matching of the overall scale with the notation of Eq.~\eqref{eq:EFT} is given by
\be
	C_S = - C_1 - 3 C_3~, \qquad
	C_T = C_1 - C_3~. \qquad
	\label{eq:ScalLQmatch}
\ee
The relation to the various observables used in the fit can be found in Appendix~\ref{app:obs}. The leading contributions to the flavour observables in Table~\ref{tab:FlavourFit} are 
\be\begin{split}
	R_{D^(*)}^{\tau/\ell} \approx &~ 1 + 2 (C_1 - C_3) + 2 (C_1 \beta_{1,s\tau} - C_3 \beta_{3,s\tau}) \frac{V_{cs}}{V_{cb}}~, \\
	\Delta C_9 = - \Delta C_{10} = &~ \frac{4 \pi}{\alpha V_{tb} V_{ts}} C_3 \beta_{s\mu} \beta_{b\mu}~,\\
	R_{b\to c}^{\mu/e} \approx &~ 1 + 2 (C_1 - C_3) \beta_{b\mu} \left( \beta_{b\mu} + \beta_{s\mu} \frac{V_{cs}}{V_{cb}}\right)~, \\
	B_{K^* \nu \nu} - 1 \propto&~ (C_1 \beta_{1,s\tau} + C_3 \beta_{3,s\tau}) ~,
\end{split}\ee
while the contributions to the radiatively generated ones can be derived simply using Eq.~\eqref{eq:ScalLQmatch}. The results of the fit of semi-leptonic flavour observables, as well as radiatively generated contributions to $Z\to\tau\bar\tau,\nu\bar\nu$ and $\tau$ decays, are illustrated in Figure~\ref{Fig:SLQ}.

A good fit can be obtained for $C_1 \sim C_3$ (to pass the limits from $\tau$ LFU decays, which are proportional to $C_T$), $\beta_{1,s\tau} \sim - \beta_{3,s\tau} \sim (\text{few}) \times V_{cb} > 0$ (to pass $B_{K^* \nu\bar\nu}$ and fit $R_{D^{*}}$), and $\beta_{s\mu} \beta_{b\mu} > 0$ (to fit $\Delta C^\mu_9$).
In particular, in this limit the leading contributions to $B_{K^* \nu \nu}$ and $\tau$ LFU observables vanish. However, radiative corrections to $Z \to \tau \bar\tau, \nu \bar\nu$ observables are enhanced by the factor of $3$ in Eq.~\eqref{eq:ScalLQmatch}, which in turn forces the size of $C_{1,3}$ to be smaller than what expected from the EFT fit, implying a $\sim 1.5 \sigma$ tension in $R_{D^(*)}$ (since we fix an upper limit on the size of $\beta_{1(3), s\tau}$). Allowing a cancellation of the radiative corrections to $Z$ couplings with a very mild tuning (at the $\sim 30\%$ level), for example due to some genuine UV contributions, the tension disappears and all flavour anomalies can be fitted at the same time.
Pure four-quark and four-lepton operators are instead generated at the one-loop level and turn out to be negligible. The greatest virtue of this scenario is the natural 
absence of significant constraints from $\Delta F=2$ processes due to the smallness of the corresponding (finite) loop amplitudes (see for example Figure~3 of Ref.~\cite{Hiller:2017bzc}).

\begin{figure}[tbp]
  \centering
      \includegraphics[width=0.6\textwidth]{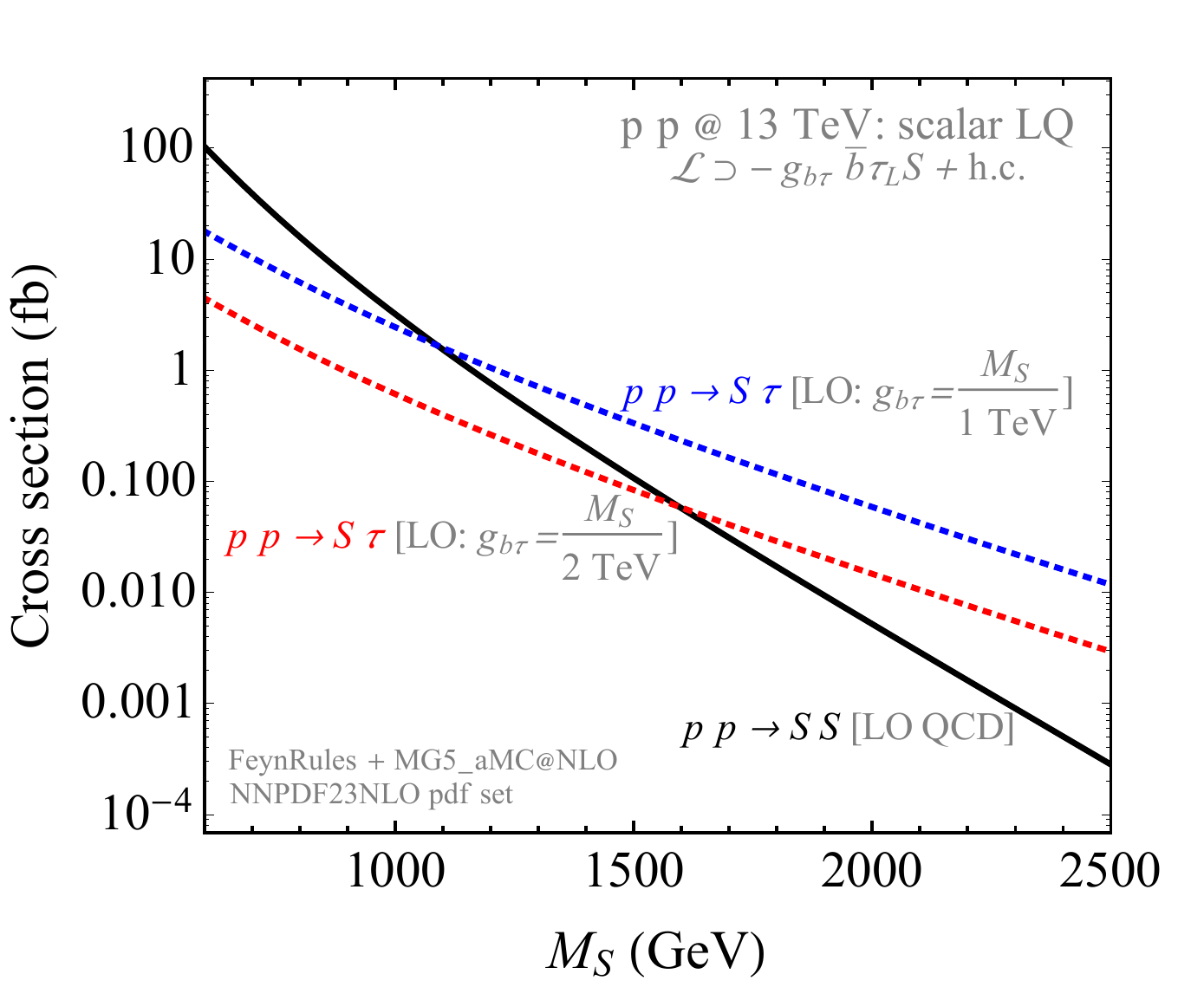}
    \caption{ \small  {Cross section (in fb) at 13~TeV $p p$ collider for: (a) scalar LQ pair production (solid black line), and (b) single LQ + $\tau$ production for the two coupling benchmarks motivated by the fit to low-energy data (dashed blue and red lines). }
    \label{Fig:ColliderLQ} }
\end{figure}

{Let us finally comment on the importance of single LQ + lepton production process in {high-$p_T$ LHC searches.} 
For illustrative purposes, we implement in {\tt FeynRules}~\cite{Alloul:2013bka} the scalar LQ field $S$ with the coupling $\mathcal L \supset - g_{b \tau}~ \bar b_R \tau_L ~S$~+h.c.~. We use {\tt MadGraph5\_aMC@NLO}~\cite{Alwall:2014hca} with the {\tt NNPDF2.3}~\cite{Ball:2012cx} NLO PDF set, to calculate the relevant cross sections at LO in QCD in the 5-flavor scheme. The results are shown in Figure~\ref{Fig:ColliderLQ}, where the solid black line is the QCD-induced LQ pair production cross section as a function of the LQ mass $M_S$. Pair production is (to a good approximation) insensitive on the LQ-$b$-$\tau$ coupling, unlike the single LQ + $\tau$ production ($g b \to S \tau$ at the partonic level). By fitting the $B$-physics anomalies, this coupling is essentially fixed {for a given value of} the LQ mass, so the cross section for $p p \to S \tau$ can be predicted in terms of the LQ mass only. Shown in dashed blue and red lines are representative examples favoured by the low-energy data, $g_{b\tau} = M_S / 1$~TeV and $g_{b\tau} = M_S / 2$~TeV, respectively. Clearly, for LQ mass $\gtrsim1$~TeV, single LQ + $\tau$ becomes an important production mechanism at the LHC.
}

\subsection{Scenario III: Colour-less Vectors}
\label{sect:vectors}

In this section, generalising the model in Ref.~\cite{Greljo:2015mma}, we assume that the effective operators in Eq.~\eqref{eq:EFT} are obtained by integrating out heavy colour-less triplet, $W^{\prime}_\mu \equiv ({\bf 1}, {\bf 3}, 0)$, and singlet, $B^\prime_\mu \equiv ({\bf 1}, {\bf 1}, 0)$, vector resonances, coupled respectively to the SM fermion triplet and singlet currents (see \cite{Greljo:2015mma} for the details on the model Lagrangian).
The effective Lagrangian obtained by integrating out these fields at the tree-level includes a set of four-fermion operators, given by
\be
  \Delta \cLT_{4f} =  - \frac{2}{v^2} J_\mu^a J_\mu^a~, \qquad\qquad  \Delta \cLS_{4f} =  - \frac{2}{v^2} J_\mu^0 J_\mu^0~,
  \label{eq:DeltaL4f}
\ee
where $J_\mu^a$ ($J_\mu^0$) is a fermion current  transforming as an $SU(2)_L$-triplet (singlet), built in terms of the SM quarks and lepton fields,
\ba
	J^a_\mu &=&  \epsilon_q \llq_{ij} \left(\bar Q_L^i \gamma_\mu T^a Q_L^j\right) 
+ \epsilon_\ell \lle_{ij} \left(\bar L_L^i \gamma_\mu T^a L_L^j\right)~,\\
	J_\mu^0 &=&\frac{1}{2} \epsilon_q^0 \llq_{ij} \left(\bar Q_L^i \gamma_\mu Q_L^j\right) 
+ \frac{1}{2} \epsilon_\ell^0 \lle_{ij} \left(\bar L_L^i \gamma_\mu L_L^j\right)~,
\label{eq:LVf}
\ea
where  $\lambda^{q,\ell}$ are Hermitian flavour matrices, $T^a\equiv\sigma^a/2$, and,
in order to be consistent with the notation of Ref.~\cite{Greljo:2015mma}, 
we included the dependence on the vector's mass in the definition of the $\epsilon_{q,\ell}^{(0)}$ parameters:
\be
	\epsilon_{q,\ell}^{(0)} = \frac{g_{\ell,q}^{(0)} m_W}{g m_V^{(0)}}~ 
	 \quad \rightarrow \quad
	 C_T = \epsilon_q \epsilon_\ell~, \quad
	C_S = \epsilon_q^0 \epsilon_\ell^0~.
\ee
For simplicity, we assume the flavour structure of the triplet and singlet currents to be the same.

In addition to semi-leptonic operators, 
this model generates tree-level contributions also to four-quark and four-lepton operators 
(see Appendix~\ref{app:ObsVect} for the details).
Among them, a particularly relevant constraint is set by the  
$\Delta F = 2$ operators contributing to $B_{s(d)}$-$\overline{B}_{s(d)}$ and $D_0$-$\overline{D}_0$ mixing,
for which we find
\be\begin{split}
	\Delta \cA_{\Delta B = 2} & \approx 154 \frac{(\llq_{bs})^2}{|V_{ts}|^2} \left( \epsilon_q^2  + (\epsilon_q^0)^2 \right) = 0.07 \pm 0.09~, \\
	\Delta \cA_{\Delta C = 2} & \approx 1.8 \left( 1+ 2 \frac{\llq_{sb}}{|V_{ts}|} \right)^2  (\epsilon_q^2 + (\epsilon_q^0)^2)   =    0.0 \pm 0.6   ~.
\end{split}
\ee
The values of $\llq_{bs} \sim 3 |V_{ts}|$ and $\epsilon_{q}^{(0)} \gtrsim 0.1$, preferred by the EFT fit of Section~\ref{sect:SLfit}, would generate contributions to $\Delta B = 2$ and $\Delta C=2$ amplitudes larger than the experimental limits
by a factor of $\sim 500$ and $\sim 20$, respectively. 
When taken at face value, these observables exclude this scenario as a viable explanation of the flavour anomalies.

A possible way out consists in introducing a coupling to right-handed up- and down-quark currents in the singlet current of Eq.~\eqref{eq:LVf}. By tuning these new couplings it is possible to completely evade the bounds from $\Delta F = 2$ processes (see Eq.~\eqref{eq:BsMixEq} in the case of $B_s$ mixing). The price to pay, however, is a tuning of the model parameters at the $\sim 10^{-4}$ level, dependent on the precise value of the hadronic matrix elements of left- and right-handed currents. We believe that such a scenario is extremely unlikely to be realised in Nature and for this reason will not pursue this further.

As anticipated in Section~\ref{sect:beyondEFT}, an alternative way in which the model could survive is to abandon the 
large $\llq_{sb}$ region selected by the EFT fit and move to the  small $\llq_{sb}$ region, 
where $\llq_{sb} = \mathcal{O}(10^{-1})\times|V_{cb}|$.
This region of parameter space was indeed the one found by the original fit of Ref.~\cite{Greljo:2015mma}, and is potentially accessible in this model adding extra Higgs-current terms in Eq.~\eqref{eq:LVf}. 
These terms are allowed by the symmetry and are naturally expected in a model of this type,
\be
	\Delta J^a_\mu = \frac{1}{2} \epsilon_H \left( i H^\dagger \stackrel{\leftrightarrow}{D^a}_\mu H \right)~, \qquad\qquad
	\Delta J_\mu^0 = \frac{1}{2} \epsilon_H^0 \left(  i H^\dagger \stackrel{\leftrightarrow}{D}_\mu H \right)~,
\label{eq:LVfHiggs}
\ee
where $H^\dagger \stackrel{\leftrightarrow}{D^a_\mu} H \equiv H^\dagger \sigma^a (D_\mu H) - (D_\mu H)^\dagger \sigma^a H$.
The effective Lagrangian at the scale $\Lambda$ becomes
\be\begin{split}
  \Delta\cL_{\rm eff} &=
  	\Delta \cLT_{4f} + \cLS_{4f}  - \frac{1}{v^2} \epsilon_H \left( i H^\dagger  \stackrel{\leftrightarrow}{D^a_\mu} H \right)  \left[ \epsilon_q \llq_{ij} \left(\bar Q_L^i \gamma_\mu \sigma^a Q_L^j\right)  + \epsilon_\ell \lle_{ij} \left(\bar L_L^i \gamma_\mu \sigma^a L_L^j\right) \right] \\
	& - \frac{1}{v^2} \epsilon_H^0 \left( i H^\dagger \stackrel{\leftrightarrow}{D}_\mu H \right)  \left[ \epsilon_q^0 \llq_{ij} \left(\bar Q_L^i \gamma_\mu Q_L^j\right) + \epsilon_\ell^0 \lle_{ij} \left(\bar L_L^i \gamma_\mu L_L^j\right) \right] \\
	& + \frac{1}{2v^2} (\epsilon_H)^2 \left(H^\dagger  \stackrel{\leftrightarrow}{D^a_\mu} H \right)^2 + \frac{1}{2v^2} (\epsilon_H^0)^2 \left(H^\dagger  \stackrel{\leftrightarrow}{D}_\mu H \right)^2 ~.
  \label{eq:DeltaLVect}
\end{split}\ee
To constrain the  free parameters appearing in this Lagrangian we take into account a series of additional observables, 
being modified at the tree-level and beyond (see Appendix~\ref{app:ObsVect} for details).
In particular, in addition to the observables already considered, here we include also 
deviations to the $Z b_L b_L$ coupling, Eq.~\eqref{eq:ZWcouplVect}, as well as to the electroweak $T$ parameter, 
Eq.~\eqref{eq:limTh}.

Performing a global fit of all the flavour and electroweak observables relevant to this model we find good solutions, capable of fitting the flavour anomalies with only a mild tuning (not exceeding the $10\%$ level) in order to evade the electroweak bounds. Given the large number of parameters of the model, we do not present plots for this case but we report here a typical benchmark point
(with $\lle_{\tau\mu} = 0$):
\be
\begin{array}{l l l l}
	\epsilon_\ell \approx 0.2 ~,\quad& \epsilon_q \approx 0.5 ~,\quad& \epsilon_H \approx -0.01~,
	\quad& \llq_{sb}/ |V_{cb}| \approx -0.07~,
	\\
	\epsilon_\ell^0 \approx 0.1 ~,\quad& \epsilon_q^0 \approx -0.1 ~,\quad& \epsilon_H^0 \approx -0.03~, 
	&\lambda^\ell_{\mu\mu} \approx 0.2 ~.
	\end{array}
\ee
corresponding to $C_S \approx - 0.01$, $C_T \approx 0.1$. This point gives a slightly lower $R_{D^{(*)}}^{\tau\ell} \approx 1.17$, while $\Delta C_9^\mu = - \Delta C_{10}^\mu \approx - 0.55$.
In this benchmark point, a value of $\lle_{\tau\mu} \lesssim 0.1$ would be compatible with the constraints 
from LFV in $\tau$ decays, Eq.~\eqref{eq:LFVtau2}, without affecting sensibly any other observable.

The only serious problem of this scenario, already encountered in Ref.~\cite{Greljo:2015mma}, is the fact that  
the large values of $\epsilon_{\ell,q}$ imply a low mass scale and large coupling of the neutral
triplet vector resonance to $b_L b_L$ and $\tau_L \tau_L$ (the singlet state can instead be heavier). 
Therefore, very stringent limits from high-$p_T$ di-tau searches apply~\cite{Faroughy:2016osc}.
As pointed out in~\cite{Greljo:2015mma,Faroughy:2016osc}, these bounds can be avoided only if the resonances
have a width significantly larger than what computed with the currents in Eq.~\eqref{eq:LVf} and~\eqref{eq:LVfHiggs}.

\section{A possible composite UV completion}
\label{sec:UV}

The mass scale of New Physics pointed out by the flavour anomalies, $M \sim$ TeV, is precisely in the ballpark of energies where New Physics  is expected to appear in order to solve the naturalness problem of the electroweak scale. It is therefore compelling to speculate on possible links between the $B$-physics anomalies and the stabilisation of the SM Higgs sector.

An interesting and wide class of SM extensions which address the hierarchy problem are the so-called composite-Higgs models. 
In this framework, the Higgs doublet is a composite pseudo-Nambu Goldstone boson (pNGB) that arises from the spontaneous breaking of a global symmetry by a new strongly-coupled sector at the TeV scale.
A UV complete description, in four spacetime dimensions, of such a setup can be realised by introducing a set of new vector-like \emph{hyper-quarks}, $\Psi_{HC}^i$, charged under a new asymptotically free gauge group, the \emph{hyper-colour} $\mathcal{G}_{HC}$, which confines at a scale $\Lambda_{HC} \sim$~(few)~TeV.
In order to include the Higgs as a pNGB, the fundamental $\Psi_{HC}^i$ should carry $\mathcal{G}_{\rm EW} = SU(2)_L \times U(1)_Y$ charges. Such models have been widely studied in the literature and offer a very rich phenomenology (see e.g Refs.~\cite{Kaplan:1983fs,Georgi:1984af,Ferretti:2013kya,Ferretti:2014qta,Cacciapaglia:2014uja,Vecchi:2015fma,Ma:2015gra,Ferretti:2016upr,Sannino:2016sfx,Agugliaro:2016clv} and references therein).
Since $\mathcal{G}_{\rm EW}$ is necessarily also a global symmetry of the strong sector, the conserved current associated to it can excite composite vector resonances with the quantum numbers of the $W^\prime$ and $B^\prime$ mediators.\footnote{More precisely, in order to satisfy LEP limits, the composite sector should enjoy the larger custodial symmetry $SU(2)_L \times SU(2)_R$. In this case the $B^\prime$ vector is accompanied by a charged $W^\prime_R$, singlet under $SU(2)_L$.}

Starting from such a scenario it is fairly natural to speculate that some of the hyper-quarks could also carry colour. In this case, one expects in general both pNGBs and heavy vectors charged under the whole SM group. By opportunely choosing the SM representations of such hyper-quarks, it is easy to obtain the scalar leptoquarks $S_1$ and $S_3$ as pNGB, or the vector leptoquarks $U_1$ and $U_3$ as composite vectors.

Such a framework, albeit without the inclusion of the Higgs as pNGB, was presented in Ref.~\cite{Buttazzo:2016kid}. In particular, the hyper-colour gauge group was fixed to $\mathcal{G}_{HC} = SU(N_{HC})$, while the vector-like hyperquarks, assumed for simplicity to be in the fundamental of $\mathcal{G}_{HC}$, were taken to be $\Psi_Q = ({\bf  N_{HC}}, {\bf 3}, {\bf 2}, Y_Q)$ and $\Psi_L = ({\bf  N_{HC}}, {\bf 1}, {\bf 2}, Y_L)$.
The condensate $\langle \bar{\Psi}^i_{HC} \Psi^j_{HC}\rangle = - f^2 B_0 \delta^{ij}$ breaks spontaneously the global chiral symmetry to the vectorial subgroup, in this case $SU(8)_L \times SU(8)_R \to SU(8)_V$.
A rich spectrum of pNGB and composite vectors arises as a consequence, containing in particular the colour-less vectors $W^\prime$ and $B^\prime$, as well as vector leptoquarks $U_1$ and $U_3$ with hypercharge $Y_{U_{1,3}} = Y_Q - Y_L$, and scalar pNGB leptoquarks $S_1$ and $S_3$ with hypercharge $Y_{S_{1,3}} = - Y_{U_{1,3}} = Y_L - Y_Q$.
By choosing $Y_{Q,L}$ one can therefore have as composite states either the vector or the scalar leptoquarks that can solve the flavour anomalies, but not both sets of mediators at the same time.

The inclusion of the Higgs as a composite pNGB in a scenario similar to that of \cite{Buttazzo:2016kid} is fairly straightforward, but it necessarily requires an 
enlargement of  the global symmetry group.
A simple example is the custodially-symmetric setup
\be\begin{aligned}
	\Psi_Q &= ({\bf  N_{HC}}, {\bf 3}, {\bf 2}, Y_Q)~, &
	\Psi_L &= ({\bf  N_{HC}}, {\bf 1}, {\bf 2}, Y_L)~, \\
	\Psi_E &= ({\bf  N_{HC}}, {\bf 1}, {\bf 1}, Y_L-1/2)~, &
	\Psi_N &= ({\bf  N_{HC}}, {\bf 1}, {\bf 1}, Y_L+1/2)~.
\end{aligned}\ee
Below the confinement scale, the spectrum of such a model would include, among many other states, two Higgs doublets and leptoquarks $S_{1,3}$ as pNGB:
\be\begin{aligned}
	H_1 &\sim (\bar{\Psi}_L \Psi_N)~, &
	H_2^c &\sim (\bar{\Psi}_L \Psi_E)~, \\
	S_{1} &\sim (\bar{\Psi}_Q \Psi_L)~, &
	S_{3}^a &\sim (\bar{\Psi}_Q \sigma^a \Psi_L)~,
\end{aligned}\ee
plus composite colour-less vectors $W^\prime$ and $B^\prime$, vector leptoquarks $U_{1,3}$, and a vector colour-octet $V$:
\be\begin{aligned}
	W_{1,\mu}^a &\sim (\bar{\Psi}_L \gamma_\mu \sigma^a \Psi_L) ~, &
	W_{2,\mu}^a &\sim (\bar{\Psi}_Q \gamma_\mu \sigma^a \Psi_Q) ~, \\
	B_{I,\mu} &\sim (\bar{\Psi}_I \gamma_\mu \Psi_I) ~, &
	V_\mu^A &\sim (\bar{\Psi}_Q \gamma_\mu \lambda^A \Psi_Q) ~, \\
	U_{1,\mu} &\sim (\bar{\Psi}_L  \gamma_\mu \Psi_Q)~, &
	U_{3,\mu}^a &\sim (\bar{\Psi}_L  \gamma_\mu \sigma^a \Psi_Q)~.
\end{aligned}\ee
For $Y_Q - Y_L = 2/3$ ($-1/3$) the vector (scalar) leptoquarks have the correct hypercharge to be the mediators of the flavour anomalies.

Despite the apparent simplicity in generating the required spectrum of mediators in this class of models, 
the construction of a complete UV 
framework is far from being trivial. 
On the one hand, the strong constraints from LEP and Higgs couplings set a lower limit on the scale $f$ close to 1~TeV. This naturally brings the composite 
vectors to a scale $M_{V}$ of several TeV, which would imply large overall couplings to (third-generation) SM fermions to fit the anomaly. 
In this respect, the scalar LQ mediators have a 
clear advantage since, being pNGB, could be much lighter ($m_S \sim 1$ TeV). 
On the other hand, a key issue to be addressed  is the coupling of the composite states to the SM fermions, with the flavour structure discussed in Section~\ref{sec:EFT}. 
In Ref.~\cite{Buttazzo:2016kid} this was obtained by a linear mixing (partial compositeness) between the composite hyper-baryons and the SM fermions, and a strong coupling of the flavour mediators with such baryons. This realisation via mixing, however, is strongly disfavoured by our present fit, suggesting a different origin of such a coupling.
A detailed study of this issue, as well as of the Higgs potential and the complete phenomenology in this class of models, is beyond the purpose of the present work.

\section{Conclusions}
\label{sec:conc}

Our analysis clearly demonstrates that a combined explanation of both charged- and neutral-current $B$-physics anomalies, 
consistent with the absence of deviations from the Standard Model so far observed in other low- and high-$p_T$ observables,  
is possible and does not require unnatural tunings in model space. The two main hypotheses we invoke in order to obtain a natural solution of the  $B$-physics anomalies are:  (i) leading NP effects in semi-leptonic operators built from the left-handed quark and lepton doublets and (ii) dominant couplings to third generation SM fermions with subleading terms for the light generations controlled by a minimally broken $U(2)_q \times U(2)_\ell$ flavour symmetry. 
As shown in Section~\ref{sect:SLfit}, a global fit to all relevant low-energy observables (including radiatively generated terms)
using an EFT based on these hypotheses leads to a  good fit to all available data, without tuned cancellations
and in terms of a small number of  free parameters (4 or 5, depending on the set of observables considered).
The preferred EFT solution, whose detailed features are listed in Section~\ref{sect:SLfit},
differs from similar analyses performed in the previous literature for two main aspects: (i) a sizeable heavy-light mixing in the quark 
sector (large $\llq_{bs}$) that, despite being consistent with the minimal breaking of the flavour symmetry, 
helps to increase the effective scale of NP and (ii) a flavour-mixing structure different from the ``pure mixing" 
scenario (i.e.~complete alignment of NP along a well-defined direction in flavour space).
Two unambiguous low-energy signatures of this EFT construction are: (i) a huge enhancement 
(of two orders of magnitude or more) of  FCNC transitions of the type $b\to s \tau\bar\tau$
(as also pointed out recently in Ref.~\cite{Crivellin:2017zlb}); (ii) the quark-flavour universality of the LFU ratios in charged 
currents, $R_{b\to u}^{\tau \ell} =R_{b\to c}^{\tau \ell}$, independently of initial- and final-state hadrons.

As discussed in Section~\ref{sect:models}, this EFT solution to the anomalies can be realised in terms of different simplified models.  
A key requirement is the absence of tree-level contributions to  $\Delta F=2$ amplitudes, naturally pointing to leptoquarks as leading mediators. 
Among them, the $SU(2)_L$-singlet vector leptoquark proposed in Ref.~\cite{Barbieri:2015yvd} stands out as an excellent candidate.
This model, other than being minimal in both the number of mediators and of free parameters,  automatically presents some of the features suggested by the more general EFT fit, 
such as the relation $C_S=C_T$ and the absence of a flavour-blind contraction among light fermions.
Unlike Ref.~\cite{Faroughy:2016osc}, we find that the model can easily escape present and near-future LHC searches for 
$\tau \bar \tau$ (and third generation leptoquarks) as a consequence of the larger new physics scale implied by the low-energy fit, solving at the same time the two most pressing problems pointed out recently in the literature~\cite{Feruglio:2016gvd,Faroughy:2016osc}.

Simplified models with a pair of scalar leptoquarks in the singlet and triplet representations of the $SU(2)_L$ gauge group emerge as a natural UV alternative to recover the same low-energy EFT. 
We find that also this setup provides an overall good description of data, albeit with a larger number of free parameters.
The main advantage of this model is that the loop contribution to $\Delta F = 2$ processes is calculable and small.

A significantly different scenario, deviating from the paradigm emerging from the EFT fit, 
is the case of a small heavy-light mixing in the quark sector (small $\llq_{sb}$)  together with a lower effective NP scale (large $C_{S,T}$), allowing possible tree-level contributions to $\Delta F=2$ amplitudes while still being compatible with the bounds.
This case is illustrated in Section~\ref{sect:vectors} by means of a simplified model with colour-less triplet and 
singlet vectors $(W^\prime, B^\prime)$.
In this case one needs some degree of model-building effort in order to cope with the constraints form electroweak 
and purely leptonic observables, with a further increase in the number of free parameters.
Besides this ``aestetic" problem, we show that an overall good description of 
low-energy data, with only a mild tuning of the free parameters, can be achieved.  
The most serious problem of this scenario, already pointed out in Ref.~\cite{Greljo:2015mma}, is the
need of very large widths (hence extra decay channels),  or very small masses,
of the vector mediators in order to pass the constraints from direct searches. 

A possible UV completion for these simplified models can be realised in the context of composite Higgs models
based on vector-like confinement: 
the mediators of the flavour anomalies could arise as composite states of a new strongly coupled sector confining at the TeV scale, of which the Higgs doublet is one of the pseudo-Nambu-Goldstone bosons. 
This class of models offers an interesting laboratory for building a more complete NP frameworks, establishing connections between the flavour anomalies and a possible solution of the electroweak hierarchy problem.

We finally stress that the flavour symmetry and symmetry-breaking structure assumed here, which by no means can be considered as exhaustive of all possible NP scenarios, 
naturally points to a connection between these  anomalies and the origin of the flavour hierarchies observed in quark and lepton mass matrices.

The large amount of data still to be collected and analysed by the flavour and high-$p_T$ LHC experiments, 
as well as from future $B$ factories, will certainly shed more light on the origin of the $B$-physics anomalies.
Should both neutral- and charged-current anomalies be confirmed as clear evidences of New Physics, 
the correlations with other low- and high-energy observables analysed in the present work will help to clarify how to extend the Standard Model in order to describe this interesting phenomenon.

\subsection*{Acknowledgements}

We thank Andrea Pattori for many useful discussions.
This research was supported in part by the Swiss National Science Foundation (SNF) under contract 200021-159720.


\appendix

\section{Flavour structure}
\label{app:flavour}

According to the hypotheses listed at the beginning of Section~\ref{sec:EFT} 
the two semi-leptonic gauge-invariant operators affected by NP are
\be
\Lambda^S_{ij, \alpha\beta } ~ (\bar Q_L^i \gamma^\mu  Q_L^j)  (\bar L_L^\alpha \gamma_\mu  L_L^\beta)~, \qquad
\Lambda^T_{ij, \alpha\beta } ~ (\bar Q_L^i \gamma^\mu \sigma^a Q_L^j)  (\bar L_L^\alpha \gamma_\mu \sigma_a L_L^\beta)~,
\label{eq:bilin}
\ee
where $\Lambda^{S(T)}_{ij, \alpha\beta }$ denote tensor structures in flavour space.   
The $U(2)_q\times U(2)_\ell$ flavour symmetry and symmetry-breaking structure implies the following decomposition for each flavour tensor
\ba
&&\Lambda_{ij, \alpha\beta } = \sum_{A,B = 0,q,\ell,q\ell}  c_{AB}  (\Gamma^A)_{i  \alpha }  (\Gamma^{B\dagger})_{j \beta}~, \qquad  c_{AB} = c_{BA}^*   \\
&& (\Gamma^0)_{i  \alpha }  = \delta_{i 3} \delta_{\alpha 3}~, \quad  
(\Gamma^q)_{i  \alpha }  = (V_q)_i \delta_{\alpha 3}~, \quad  
(\Gamma^\ell)_{i  \alpha }  = \delta_{i 3}  (V_\ell)_{\alpha}~, \quad  
(\Gamma^{q\ell})_{i  \alpha }  = (V_q)_i   (V_\ell)_{\alpha}~, \nonumber
\ea 
where  $c_{AB}$ are in general $\mathcal{O}(1)$ parameters and we have neglected the possible flavour-blind contractions of the $U(2)_{q,\ell}$ doublets. 
Without loss of generality one can set $c_{00}=1$. Each tensor structure then contains 15 free real parameters (the 6 complex $c_{A\not=B}$ and the 3 remaining $c_{AA}$),
that reduce to 9 in the limit of CP conservation (CPC). 

A significant reduction of the independent parameters occurs under the hypothesis that the semi-leptonic bilinears in (\ref{eq:bilin}) are obtained by the contraction of two currents: either two LQ currents or two colour-less currents (LL$\times$QQ).
Omitting $SU(2)_L$ indices, the general structure of these currents is 
\ba 
J_{LQ}^\mu &=& \bar Q^i \gamma^\mu L^\beta \left[  \delta_{i 3} \delta_{\beta 3} + \alpha_q^*  ~(V_q)_i \delta_{\beta 3}  + \alpha_\ell ~ \delta_{i 3}  (V^*_\ell)_{\beta} + \delta ~
 (V_q)_i (V^*_\ell)_{\beta}   \right] \equiv \beta_{i\beta} ~ \bar Q^i \gamma^\mu L^\beta~,  \no \\
J_{LL}^\mu &=& \bar L^\alpha \gamma^\mu L^\beta \left[  \delta_{\alpha 3} \delta_{\beta 3} + a_\ell ~ \delta_{\alpha 3} (V^*_\ell)_\beta  +
a^*_\ell ~ (V_\ell)_\alpha  \delta_{\beta 3}  + b_\ell  ~ (V_\ell)_\alpha   (V^*_\ell)_\beta  \right] \equiv   \lle_{\alpha \beta}~\bar L^\alpha \gamma^\mu L^\beta~,  \quad \no \\
J_{QQ}^\mu &=& \bar Q^i \gamma^\mu Q^j \left[  \delta_{i 3} \delta_{j 3} + a_q ~ \delta_{i 3} (V^*_q)_j  +
a^*_q ~(V_q)_i \delta_{j3}  + b_q ~(V_q)_i  (V^*_q)_j  \right]~ \equiv \llq_{ij} ~\bar Q^i \gamma^\mu Q^j~,
\label{eq:currents}
\ea
and the corresponding tensor coefficients obtained in the two cases are reported in Table~\ref{tab:tensors}.

The most constrained case is the LQ  scenario, which is described by 3 complex parameters ($\alpha_{q,\ell}$ and $\delta$)
that reduce  to 3 real parameters in the CPC limit. Interestingly enough, in such case the flavour-blind contractions of the $U(2)_{q,\ell}$ doublets  are  automatically forbidden. 

In the LL$\times$QQ case we have 2 complex ($a_{q,\ell}$) 
and 2 real ($b_{q,\ell}$)  parameters, that reduce to 4 real parameters in the CPC limit. 
In this case flavour-blind contractions of the $U(2)_{q,\ell}$ doublets are not automatically forbidden: their absence must be imposed as an additional dynamical condition.

\begin{table}[t]
\begin{center}
\begin{tabular}{c||c|c|c|c}
  $J_{LQ}^{\phantom{\dagger}} J_{LQ}^\dagger$	&	$\Gamma^0$ & $\Gamma^q$ & $\Gamma^\ell$ & $\Gamma^{q\ell}$ \\  \hline\hline
$\Gamma^{0~~}$	& 	$1$ & $\alpha_q^*$ & $\alpha_\ell^*$ & $\alpha_q^* \alpha_\ell^*$ \\ \hline
$\Gamma^{q\dagger~}$	& $\alpha_q$ 	& $|\alpha_q^*|^2$ & $\delta^*$  & $\alpha_q \delta^*$  \\ \hline
$\Gamma^{\ell\dagger~}$ 	&    $\alpha_\ell$   	&  $\delta$ & $|\alpha_\ell^*|^2$ &   $\alpha_\ell^*  \delta$  \\  \hline
$\Gamma^{q\ell\dagger}$ 	&   $\alpha_q \alpha_\ell$   	&  $\alpha_q^* \delta$   & $\alpha_\ell \delta^*$  &  $|\delta|^2$  \\
\end{tabular}  \qquad\qquad 
\begin{tabular}{c||c|c|c|c}
    $J_{QQ} J_{LL}$	& $\Gamma^0$ & $\Gamma^q$ & $\Gamma^\ell$ & $\Gamma^{q\ell}$ \\  \hline\hline
$\Gamma^{0~~}$	& 	$1$ & $a^*_q$ & $a^*_\ell  $ & $a_\ell^* a^*_q$ \\ \hline
$\Gamma^{q\dagger~}$	& $a_q$	& $b_q $ &  $a_\ell^* a_q$  & $a_\ell^* b_q $  \\ \hline
$\Gamma^{\ell\dagger~}$	&    $a_\ell  $    	& $a_\ell a^*_q$  & $b_\ell$ &   $b_\ell a_q^*$  \\  \hline
$\Gamma^{q\ell\dagger}$ 	&  $a_\ell a_q$    &   $a_\ell  b_q$ &  $b_\ell a_q$   &  $ b_\ell b_q$  \\
\end{tabular}
\end{center}
\caption{Tensor structures in the LQ and   QQ$\times$LL	 case.
\label{tab:tensors}}
\end{table}

The two scenarios are not equivalent as long as we consider terms with more than one spurion, but in both cases we can assume as free parameters the set $\{\llq_{bs}, \lle_{\tau\mu}, \lle_{\mu\mu} \}$, defined by 
\ba
\Lambda_{bs,\tau\tau}  \equiv  \llq_{bs}~, \qquad \Lambda_{bb,\tau\mu}  \equiv  \lle_{\tau\mu}~,
\qquad  \Lambda_{bs,\mu\mu} \equiv \llq_{bs} \lle_{\mu\mu}~, 
\ea
which can be expressed as
\be
\begin{array}{llllll}
J_{QQ} J_{LL} & \to & 	\llq_{bs} = a_q V_{ts}~,   & \lle_{\tau\mu} = a_\ell V_{\tau\mu}~, &   \lle_{\mu\mu} = b_\ell |V_{\tau\mu}|^2~,   \\
J_{LQ} J_{LQ}^\dagger & \to  & 	 \llq_{bs} = \alpha_q V_{ts} \equiv  \beta^*_{s\tau}~,  &   \lle_{\tau\mu} = \alpha_\ell V_{\tau\mu} \equiv  \beta_{b\mu}~,
&   \lle_{bs}\lle_{\mu\mu} =  \alpha_\ell  \delta^*  |V_{\tau\mu}|^2  \equiv   \beta_{b\mu}  \beta^*_{s\mu} ~,  \label{eq:LQbeta} \\
\end{array}
\ee
in terms of the current parameters.  Note that in both cases the relation $\Lambda_{bs,\tau\mu} = \llq_{bs} \lle_{\tau\mu}$ is 
satisfied, while the expression for $\Lambda_{bb,\mu\mu}$ in terms of $\{\llq_{bs}, \lle_{\tau\mu}, \lle_{\mu\mu} \}$
is different:
\ba
 \left. \Lambda_{bb,\mu\mu} \right|_{{\rm LL}\times{\rm QQ}} = \lle_{\mu\mu}~, \qquad    \left. \Lambda_{bb,\mu\mu} \right|_{{\rm LQ}\times{\rm QL}} 
 = |\lle_{\tau\mu}|^2 ~.  
\ea
In the LQ case, the definition of these three parameter determines completely the flavour structure of the system.
In the  QQ$\times$LL case we have one extra free parameter  that can be defined  from
$\Lambda_{sd,\tau\tau}  \equiv  \llq_{sd}$:
\ba
 \left. \llq_{sd} \right|_{{\rm LL}\times{\rm QQ}} = b_q V^*_{ts} V_{td}~({\rm free})~, \qquad    \left. \llq_{sd} \right|_{{\rm LQ}\times{\rm QL}} 
 =  |\llq_{bs}|^2 \frac{V^*_{ts} V_{td}}{ |V_{ts}|^2} ~({\rm fixed})~.  
 \label{eq:lambdasd}
\ea

\subsection{Basis alignment and pure-mixing scenario}

The identification of the $U(2)_q\times U(2)_\ell$  singlets with the 
third-generation down-type quarks and charged leptons, i.e.~the basis choice in Eq.~(\ref{eq:basis0}),  
is somehow arbitrary. On general grounds we expect 
\be
 Q^{\rm singlet}_{L} \equiv Q_L^3 + \theta_q \sum_{i=1,2}  (V^*_q)_i  Q^i_L~,  \qquad 
 L^{\rm singlet}_{L} \equiv L_L^3 + \theta_\ell \sum_{\alpha=1,2}  (V^*_\ell)_\alpha  L^\alpha_L~,  \qquad 
\label{q3}
\ee
were $\theta_{q,\ell}$  are complex $O(1)$ parameters controlling the possible basis mis-alignment.
Under the change of basis  $Q^3 \to Q_L^3 +  (V^*_q)_i  Q^i_L$ (and similarly for the leptons), 
the current parameters defined in Eq.~(\ref{eq:currents}) undergo the following transformations:
\ba
 J_{QQ} J_{LL}:  &\quad& 
a_{q(\ell)} \to  a_{q(\ell)} + \theta_{q(\ell)}~,  \qquad  b_{q(\ell)} \to   b_{q(\ell)}  +|\theta_{q(\ell)}|^2 + 2 \Re[\theta_{q(\ell)} a_{q(\ell)}]~,  \\ 
 J_{LQ} J_{LQ}^\dagger :  &\quad &  \alpha_{q(\ell)} \to  \alpha_{q(\ell)} + \theta_{q(\ell)}~,  \qquad  \delta \to  \delta  + \theta^*_{q} \theta_\ell +  \theta_\ell \alpha_q^* +\theta_q^* \alpha_\ell~.
\ea
From these transformations we deduce that the parameters $a_{q(\ell)}$, $b_{q(\ell)}$ and $\delta$ are all expected to be 
$O(1)$ unless some specific basis choice is adopted. This implies in particular
\be
\llq_{bs} = O(1) \times |V_{ts}| = O(1) \times |V_{cb}|~.
\label{eq:bsbound}
\ee

A particularly restrictive scenario, that can be implemented both in the LQ or QQ$\times$LL cases,
is the so-called pure-mixing scenario, i.e.~the hypothesis that there exists a flavour basis where the NP interaction 
is completely aligned along the flavour singlets. Under this assumption there exists a 
basis where $a_{q(\ell)}= b_{q(\ell)}$ or $\alpha_{q(\ell)}=\delta=0$ for all flavour tensors. 
This imply all flavour tensors are described by only two parameters, $\theta_{q}$ and $\theta_\ell$, 
that control the  basis mis-alignment. In both cases, in this specific limit, one arrives to the prediction
\be
\lle_{\mu\mu} = | \theta_\ell |^2  | V_{\tau\mu} |^2  > 0~.
\ee

\section{Experimental observables}
\label{app:obs}

\subsection{Minimal set relevant for the semi-leptonic operators}

\boldmath
\subsubsection*{LFU in charged-current semi-leptonic $B$ decays}
\unboldmath

From the combined HFAG fit \cite{Amhis:2016xyh} (for Moriond EW 2017), assuming $R_{D} = R_{D^{(*)}}$ (to which we add the recent LHCb measurement of $ R_{D^{(*)}}$ with hadronic $\tau$ decays \cite{LHCbRDstar}, assuming zero correlation)
one gets 
\be
	R_{b\to c}^{\tau \ell} \equiv R_{D^{(*)}} \equiv \frac{\mathcal{B}(B \to D^{(*)}\tau \nu)_{\rm exp} / \mathcal{B}(B \to D^{(*)}\tau \nu)_{\rm SM}}{\frac{1}{2} \sum_{\ell = \mu, e} \left[\mathcal{B}(B \to D^{(*)}\ell \nu)_{\rm exp} / \mathcal{B}(B \to D^{(*)}\ell \nu)_{\rm SM} \right]} = 1.237 \pm 0.053~.
\ee
The expression of this ratio in presence of a single flavour breaking structure $\lambda_{ij}^{q,\ell}$ (as we assume in the EFT and vector mediator cases) is
\be
	R_{D^{(*)}} = \frac{|1 + C_T (1 + \Delta)|^2 + |C_T \lle_{\tau\mu} (1 + r \Delta) |^2 }{\frac{1}{2} \left( |1 + C_T \lle_{\mu\mu} (r^{-1} +   \Delta)|^2 + |C_T \lle_{\tau\mu} (1 + \Delta) |^2 + 1 \right)}~,
\ee
where
\be
\Delta = \llq_{sb} \frac{V_{cs}}{V_{cb}}  +  \llq_{db} \frac{V_{cd}}{V_{cb}}  =  - \frac{V_{tb}^* }{V^*_{ts} } \llq_{sb}~,
\label{eq:Deltacb}
\ee
with $r=1$  for a colour-less mediator and $r = \beta_{s\mu} / (\beta_{b\mu} \beta_{s\tau}) = \lle_{\mu\mu}/(\lle_{\tau\mu})^2$
in the vector LQ case (where, as already discussed, $\llq_{sb} =  \beta_{s\tau}$,  $\lle_{\tau\mu} =\beta_{b\mu}$, and 
$\lle_{bs}\lle_{\mu\mu} =  \beta_{b\mu}  \beta^*_{s\mu}$).
On the r.h.s.~of  Eq.~(\ref{eq:Deltacb}) we have used  $\llq_{db}/\llq_{sb} = V_{td}^*/ V_{ts}^*$ and CKM unitarity.
Note that the value of $\Delta$ thus defined is nothing but the coefficient of the $U(2)_q$ breaking spurion in the currents which, by construction, is  flavour independent. 

Similar  LFU ratios ($\tau$ vs.~$\ell = \mu, e$) can be constructed also in $b\to u $ transitions. 
In this case the expressions are identical to those reported above
since 
\be
 \llq_{sb} \frac{V_{us}}{V_{ub}}  +  \llq_{db} \frac{V_{ud}}{V_{ub}}  = \llq_{sb} \frac{V_{cs}}{V_{cb}}  +  \llq_{db} \frac{V_{cd}}{V_{cb}}  
 \equiv
 \Delta~. 
\ee
Thanks to the flavour symmetry we therefore expect a universal enhancement of $b\to c$  and $b\to u $ transitions with $\tau$ leptons in the final state,
irrespective of the value of $\llq_{bs}$. So far, the only measurement of $b\to u \tau\nu$ transitions is 
$\mathcal{B}(B \to \tau \nu)$ that, according to the global UTFit analysis~\cite{Bona:2017cxr}, leads to 
\be
  \frac{\mathcal{B}(B \to D^{(*)}\tau \nu)_{\rm exp}}{ \mathcal{B}(B \to D^{(*)}\tau \nu)_{\rm SM} } = 1.31\pm 0.27~,
\ee
supporting the prediction $R_{b\to u}^{\tau \ell} =R_{b\to c}^{\tau \ell}$. 

Deviations from LFU in the first two generations of leptons are instead constrained by~\cite{Olive:2016xmw}
\be
	R_{b\to c}^{\mu e} \equiv \frac{\mathcal{B}(B \to D^{(*)}\mu \nu)_{\rm exp} / \mathcal{B}(B \to D^{(*)}\mu \nu)_{\rm SM}}{ \mathcal{B}(B \to D^{(*)}e \nu)_{\rm exp} / \mathcal{B}(B \to D^{(*)}e \nu)_{\rm SM}} = 1.000 \pm 0.021~.
\ee
The expression of this ratio, again under the assumption of a single flavour breaking structure, is
\ba
	R_{b\to c}^{\mu e} &=&  |1 + C_T \lle_{\mu\mu} (r^{-1} +   \Delta)|^2 + |C_T \lle_{\tau\mu} (1 + \Delta) |^2~.
\ea

In the case of the scalar LQ the expressions of the LFU ratios are slightly more involved due to 
the different flavour couplings of singlet and triplet operators. 
Neglecting CKM suppressed terms and setting 
$\beta_{3,s\mu} = \beta_{1,s\mu}$ and $\beta_{1,b\mu} = \beta_{3,b\mu}$ 
we get 
\ba
	R_{D^{(*)}} &=& \frac{|1 + C_1 (1 + \Delta_{1}) - C_3 (1 + \Delta_{3}) |^2 + |(C_1-C_3) (\beta_{b\mu} + \beta_{s\mu} V_{cs}/V_{cb}) |^2 }{\frac{1}{2} \left( |1 + (C_1-C_3) \beta_{b\mu} (\beta_{b\mu} + \beta_{s\mu} V_{cs}/V_{cb} ) |^2 + | \beta_{b\mu} [C_1 (1 +  \Delta_{1}) - 
	C_3 (1 +  \Delta_{3}) ] |^2 + 1 \right)}~,
	\label{eq:SLQRD}  \nonumber  \\
 R_{b\to c}^{\mu e}  &=&    |1 + (C_1-C_3) \beta_{b\mu} (\beta_{b\mu} + \beta_{s\mu} V_{cs}/V_{cb} ) |^2 + | \beta_{b\mu} [C_1 (1 +  \Delta_{1}) - 
	C_3 (1 +  \Delta_{3}) ] |^2 ~,  \nonumber  \\
\ea
where $\Delta_{i} =   \beta_{i,s\tau}  V_{cs}/ V_{cb}$.

\boldmath
\subsubsection*{$b \to s \ell\ell$ transitions}
\unboldmath

Many groups performed global fits of the available $b \to s \mu\bar\mu$ data, see e.g. Refs.~\cite{Capdevila:2017bsm,Altmannshofer:2017yso,DAmico:2017mtc,Ciuchini:2017mik} for the latest updates. In this work we use the results of~\cite{Capdevila:2017bsm}.
In our main setup only the left-handed fields are strongly coupled to the new physics, therefore we are interested in the direction $\Delta C_9^\mu = - \Delta C_{10}^\mu$. In this case the global fits provide
\be
	\Delta C_9^\mu = - \Delta C_{10}^\mu = - 0.61 \pm 0.12~.
\ee
The expressions of these modified Wilson coefficients in the three scenarios we have considered are 
\begin{align}
	\Delta C_9^\mu &= - \Delta C_{10}^\mu = - \frac{\pi}{\alpha V_{tb} V_{ts}^*} (C_T + C_S) \llq_{bs} \lle_{\mu\mu}, & \text{\sc\small (eft and vector resonances)}\\
	\Delta C_9^\mu &= - \Delta C_{10}^\mu = - \frac{2\pi}{\alpha V_{tb} V_{ts}^*} C_U \beta_{s\mu} \beta_{b\mu}, & \text{\sc\small (vector leptoquark)}\\
	\Delta C_9^\mu &= - \Delta C_{10}^\mu = \frac{4\pi}{\alpha V_{tb} V_{ts}^*} C_3 \beta_{s\mu} \beta_{b\mu}. & \text{\sc\small (scalar leptoquark)}
\end{align}
The results relevant to $b \to s \tau\bar\tau$ transitions are  simply obtained 
from those above replacing $\llq_{bs} \lle_{\mu\mu}$ ($\beta_{s\mu} \beta_{b\mu}$) with $\llq_{bs}$ ($\beta_{s\tau} \beta_{b\tau}$).

\boldmath
\subsubsection*{Limits from $B \to K^{(*)} \nu \nu$}
\unboldmath

The branching ratio of the rare FCNC decay $B \to K^{*} \nu \nu$ is bounded from above as~\cite{Olive:2016xmw}
\be
	B_{K^{(*)}\nu\bar\nu} \equiv \frac{\mathcal{B}(B \to K^{(*)} \nu \nu)_{\rm exp}}{\mathcal{B}(B \to K^{(*)} \nu \nu)_{\rm SM}} < 5.2\quad [ 95\% \text{CL}]~.
\ee
In our setup this ratio is potentially affected by large corrections. 
In the EFT and vector-inspired setup we find
\be\begin{split}
	B_{K^{(*)}\nu\bar\nu} &= \frac{1}{3} \left[ 1 
	+ \left| 1 + \frac{\pi}{\alpha} \frac{(C_T - C_S) \llq_{bs}}{C_\nu^{\rm SM} V_{ts}^* V_{tb}}\right|^2 
	+ \left| 1 + \frac{\pi}{\alpha} \lle_{\mu\mu} \frac{(C_T - C_S) \llq_{bs}}{C_\nu^{\rm SM} V_{ts}^* V_{tb}} \right|^2 + \right. \\
	& \left.  + 2 \left| \frac{\pi}{\alpha} \lle_{\tau\mu} \frac{(C_T - C_S) \llq_{bs}}{C_\nu^{\rm SM} V_{ts}^* V_{tb}} \right|^2 \right]~,
\end{split}\ee
with $C_\nu^{\rm SM} = -6.4$,
while in the vector leptoquark setup, where $C_T = C_S = C_U$,  all corrections to this observable vanish.
In the scalar leptoquark case, where we distinguish between $\beta_{1,s\tau}$ and $\beta_{3,s\tau}$, we get 
\be\begin{split}
	B_{K^{(*)}\nu\bar\nu} &= \frac{1}{3} \left[ 1 
	+ \left| 1 + \frac{2\pi}{\alpha} \frac{ C_1 \beta_{1,s\tau} + C_3 \beta_{3,s\tau} }{C_\nu^{\rm SM} V_{ts}^* V_{tb}} \right|^2 
	+ \left| 1 + \frac{2\pi}{\alpha} \beta_{b\mu} \frac{C_1 \beta_{1,s\tau} + C_3 \beta_{3,s\tau} }{C_\nu^{\rm SM} V_{ts}^* V_{tb}} \right|^2 + \right. \\
	& \left.  +  \left| \frac{2\pi}{\alpha} \beta_{s\mu} \frac{C_1 + C_3}{C_\nu^{\rm SM} V_{ts}^* V_{tb}} \right|^2
	+ \left| \frac{2\pi}{\alpha} \beta_{s\mu} \beta_{b\mu} \frac{C_1 + C_3}{C_\nu^{\rm SM} V_{ts}^* V_{tb}} \right|^2 \right]~.
\end{split}\ee
As pointed out in \cite{Bordone:2017lsy}, the $U(2)_q$ symmetry implies a close correlation of the NP effects in  
$\cB(B \to K^{(*)}\nu\bar\nu)$ and $\cB(K^+ \to \pi^+ \nu\bar\nu)$. Because of Eq.~(\ref{eq:lambdasd}), 
this involves no new free parameters in the vector LQ model. However, in the latter case  
the present constraint from $\cB(B \to K^{(*)}\nu\bar\nu)$ turns out to be more stringent~\cite{Bordone:2017lsy}.

\boldmath
\subsection{Radiative corrections to $ Z$ and $\tau$ observables}
\unboldmath

Here we list the set of relevant observables generated at one-loop in the leading-log approximation~\cite{Feruglio:2017rjo}. Numerical coefficients are computed assuming the matching scale $\Lambda = 2$~TeV. 
In the case of the scalar leptoquark, these effects can be included via the relation of Eq.~\eqref{eq:ScalLQmatch}.

\boldmath
\subsubsection*{Left-handed $Z \tau \tau$ and $Z \nu \nu$ couplings}
\unboldmath

One-loop correction to $Z$ couplings with the left-handed $\tau$ lepton and neutrinos due to the RG evolution of the semi-leptonic operators is \cite{Feruglio:2017rjo}
\be\begin{split}
	\delta g_{\tau_L}^Z &= \frac{1}{16\pi^2} \left( 3 y_t^2 (C_T - C_S) L_t - g^2 C_T L_z - \frac{g_1^2}{3} C_S L_z \right) \approx - 0.043 C_S + 0.033 C_T~, \\
	\delta g_{\nu L}^Z &= \frac{1}{16\pi^2} \left( 3 y_t^2 (-C_T - C_S) L_t + g^2 C_T L_z - \frac{g_1^2}{3} C_S L_z \right) \approx  - 0.043 C_S - 0.033 C_T ~, 
	\label{eq:RGZtau}
\end{split}\ee
where $L_{t,z} = \log \Lambda / m_{t,z}$ and we fixed $\Lambda = 2$~TeV and $y_t^{\overline{\rm MS}}(m_t) \approx 0.94$.
Neglecting the small correlations reported in Table~7.7 of~\cite{ALEPH:2005ab}, we find
\be
	\delta g_{\tau_L}^Z =-0.0002\pm0.0006~,
\ee
taking $s_W^2=0.23126$~\cite{Olive:2016xmw}. A modified $Z$ coupling to $\tau$-neutrino impacts the invisible $Z$ decay reported as the number of neutrinos~\cite{ALEPH:2005ab}, $N_\nu = 3 + 4 \delta g_{\nu L}^Z = 2.9840 \pm 0.0082$, providing
\be
	\delta g_{\nu L}^Z = -0.0040 \pm 0.0021~.
\ee

\boldmath
\subsubsection*{LFU in $\tau$ decays (radiative)}
\unboldmath

One-loop corrections modify $W$ couplings to $\tau$ lepton which are tested at the per-mil level in $\tau$ decays~\cite{Pich:2013lsa}.
Combining the limits on LFU ratios shown in Ref.~\cite{Pich:2013lsa} we get
\be
	|g_\tau^W / g_\mu^W| = 0.9995 \pm 0.0013~, \qquad
	|g_\tau^W / g_e^W| = 1.0030 \pm 0.0015~. \qquad
\ee
If $g_\mu^W = g_e^W \equiv g_\ell^W$, then one has
\be
	|g_\tau^W / g_\ell^W| = 1.00097 \pm 0.00098~.
\ee
Radiative corrections contribute to this ratio as \cite{Feruglio:2016gvd}
\be
	|g_\tau^W / g_\ell^W| =  1 - \frac{6 y_t^2}{16\pi^2} C_T \log \frac{\Lambda}{m_t}  \approx 1 - 0.084 C_T~,
	\label{eq:LFUrad}
\ee
where we fixed $\Lambda = 2$~TeV.

\boldmath
\subsubsection*{LFV in $\tau$ decays (radiative)}
\unboldmath

Renormalisation group effects from the semi-leptonic operators also generate LFV $Z\tau\mu$ couplings. The main effect, proportional to $y_t^2$, is given by \cite{Feruglio:2016gvd,Feruglio:2017rjo}
\be
	\delta g_{\tau\mu L}^Z = - \frac{3 y_t^2}{16 \pi^2} (C_S - C_T) \lle_{\tau\mu} \log \frac{\Lambda}{m_t}~.
\ee
At low energy, this induces LFV $\tau$ decays such as:
\be
	\frac{\mathcal{B}(\tau \to 3 \mu)}{\mathcal{B}(\tau \to \mu \nu \nu)} = \left[ 2 \left( -2 (g_{\mu L}^Z)^{\rm SM} \delta g_{\tau\mu L}^Z \right)^2 + \left( -2 (g_{\mu R}^Z)^{\rm SM} \delta g_{\tau\mu L}^Z \right)^2 \right]~,
	\label{eq:LFVtau1}
\ee
where $(g_{\mu L}^Z)^{\rm SM} = -1/2 + s_{\theta_W}^2$ and $(g_{\mu R}^Z)^{\rm SM} = s_{\theta_W}^2$. Using $\mathcal{B}(\tau \to \mu \nu \nu) \approx 17.4 \%$ and a scale $\Lambda = 2$~TeV one obtains
\be
	\mathcal{B}(\tau \to 3 \mu) \approx 2.5 \times 10^{-4} (C_S - C_T)^2 (\lle_{\tau\mu})^2 < 1.2 \times 10^{-8} ~.
	\label{eq:LFVrad}
\ee
While this is vanishing in the vector leptoquark model, in the case of the scalar leptoquarks the expression is obtained simply by substituting $C_S - C_T =2 (C_1 + C_3)$ and $\lle_{\tau\mu} = \beta_{b\mu}$.

\subsection{Connection to four-quark and four-lepton operators for the vector model}
\label{app:ObsVect}

Here we present the additional observables, and their functional dependence on the model's parameters, 
relevant to the simplified model with colour-less vectors. In particular, these are $\Delta F = 2$ processes, generated at the tree-level by pure four-quark operators, as well as tree-level contributions to electroweak and $\tau$ decays observables.

\boldmath
\subsubsection*{$\Delta F=2$ processes}
\unboldmath

The effective $\Delta F=2$ Lagrangian reads
\begin{align}
	\Delta \mathcal{L}_{\Delta B = 2} &= - 
	\frac{G_F}{\sqrt{2}} \left\{
	\left[  (\epsilon_q^2 + (\epsilon_q^0)^2)   (\llq_{ib})^2 \right] 
	 (\bar{b}_L \gamma_\mu d^i_L)^2 + (\epsilon_q^0)^2(\lambda_{ib}^d)^2 (\bar b_R \gamma_\mu d_R^i)^2\right.\nonumber\\
	  &\left. \quad + 2(\epsilon_q^0)^2 \lambda_{ib}^q\lambda_{ib}^d(\bar b_L \gamma_\mu d_L^i)(\bar b_R \gamma_\mu d_R^i)\right\}
	 + \text{h.c.}~, \label{DeltaB}\\
	\Delta \mathcal{L}_{\Delta C = 2} &= - 
	 \frac{G_F}{\sqrt{2}}  \left[ (\epsilon_q^2 + (\epsilon_q^0)^2)   \left(1 - 2 \lambda_{sb}^q\frac{V_{tb}^*}{V_{ts}^*}\right)^2 \right]
	\left( V_{ub} V_{cb}^* \right)^2 (\bar{u}_L \gamma_\mu c_L)^2 + \text{h.c.}
\end{align}
In \eqref{DeltaB} we added also a coupling to a right-handed singlet current, such that the parameter $\lambda_{sb}^q$ can be tuned to cancel the contribution arising from $\llq_{sb}$ 
in $\Delta B=2$ amplitudes.

From the global CKM fit allowing generic NP contributions to $B_{s(d)}$--$\bar B_{s(d)}$ mixing
one finds~\cite{Bona:2017cxr}
\be
\Delta \cA_{\Delta B = 2} =    \frac{\cA^{\rm SM+NP}_{\Delta B=2}}{\cA_{\Delta B=2}^{\rm SM}}-1 = 0.07 \pm 0.09~. 
\ee
Taking into account the SM contribution to the mixing amplitude~\cite{Buras:2001ra} we 
arrive to 
\be
	\Delta \cA_{B_s}^{\Delta F = 2} = 
	\frac{1}{(V_{tb}^* V_{ts})^2 R_{\rm SM}^{\rm loop}}   
	\left[\epsilon_q^2(\llq_{sb})^2 + (\epsilon_q^0)^2 \big[(\llq_{sb})^2 + (\lld_{sb})^2 - 7.14 \llq_{sb} \lld_{sb}\right]
\label{eq:BsMixEq} 
\ee
where $R_{\rm SM}^{\rm loop} = \frac{\alpha}{4 \pi s_W^2} S_0(x_t) \approx 6.5 \times 10^{-3}$.
The coupling to the right-handed current generates also a tiny contribution to  $\Delta C_{9(10)}^\prime$,
\be
\Delta C_{9}^\prime = - \Delta C_{10}^\prime = - \frac{\pi}{\alpha V_{tb} V_{ts}^*} \epsilon_q^0 \epsilon_\ell^0 \lld_{sb} \lambda^\ell_{\mu\mu}~,
\ee
which does not pose a significant constraint given the present bounds 
on these Wilson coefficients~\cite{Capdevila:2017bsm}.

In the case of $D$--$\bar D$ mixing, normalising the (magnitude) of the coefficient of the $\Delta C=2$ operator to its maximally allowed 
contribution ($\Lambda _{uc} > 3 \times 10^{3}~{\rm TeV}$~\cite{Isidori:2013ez}) we derive the constraint
\be
\Delta \cA_{\Delta C = 2}  \approx 1.8 \left( 1+ 2 \frac{\llq_{sb}}{|V_{ts}|} \right)^2  (\epsilon_q^2 + (\epsilon_q^0)^2)   =    0.0 \pm 0.6~.   
\ee

\boldmath
\subsubsection*{Electroweak, $\tau$ LFU, and $\tau$ LFV constraints}
\unboldmath

In the case of the heavy $W^\prime$ and $B^\prime$ vectors, other than the LFU-violating contributions to $Z$ and $W$ couplings to fermions due to renormalisation group effects from the semi-leptonic operators, there are other relevant effects generated by the effective Lagrangian in Eq.~\eqref{eq:DeltaLVect}:
\begin{itemize}
\item Tree-level contributions to $Zb_Lb_L$, $Z\tau_L\tau_L$, $Z\nu_\tau\nu_\tau$, and $W\tau_L \nu_\tau$ couplings from the effective operators build out of Higgs and fermion currents, as well as from the RG evolution of the semi-leptonic, four-quark, and four-lepton operators \cite{Jenkins:2013wua}. Since the numerically larger contributions are those proportional to $y_t^2$, the leading contributions are from the semi-leptonic and four-quark operators (where the quark loop is closed). Fixing the cutoff scale to $\Lambda = 1$ TeV (since in this fit $C_{S,T}$ are larger than in the EFT fit), one has:
\be\begin{split}
	\Delta g_{\nu_\tau L}^{Z} &= \frac{1}{2} (\epsilon_H^0 \epsilon_\ell^0 - \epsilon_H \epsilon_\ell ) - 0.031 \epsilon_\ell^0 \epsilon_q^0 - 0.024 \epsilon_\ell \epsilon_q ~, \\
	\Delta g_{\tau L}^{Z} &= \frac{1}{2} ( \epsilon_H^0 \epsilon_\ell^0 + \epsilon_H \epsilon_\ell ) - 0.031 \epsilon_\ell^0 \epsilon_q^0 + 0.024 \epsilon_\ell \epsilon_q~,\\
	\Delta g_{\tau}^{W} &= - \epsilon_H \epsilon_\ell - 0.060 \epsilon_\ell \epsilon_q~,  \\
	\Delta g_{b L}^{Z} &= \frac{1}{2} ( \epsilon_H^0 \epsilon_q^0 + \epsilon_H \epsilon_q ) - 0.030 (\epsilon_q^0)^2 + 0.010 (\epsilon_q)^2 = (3.3 \pm 1.6) \times 10^{-3}~.
	\label{eq:ZWcouplVect}
\end{split}\ee
\item Tree-level contributions to $\tau$ decays from four-lepton operators and modified $W\tau_L\nu_\tau$ coupling, affecting $\Gamma_{\tau\to\mu} / \Gamma_{\mu \to e}$  \cite{Pich:2013lsa}, from which one has:
\be\begin{split}
	R_\tau^{\tau/e} &= \left| 1 - \epsilon_H \epsilon_\ell - \frac{3 y_t^2}{16\pi^2} \epsilon_\ell \epsilon_q \log \frac{\Lambda^2}{m_t^2} + \epsilon_\ell^2 \lle_{\mu\mu} + \frac{1}{2} ( (\epsilon_\ell^0)^2 - \epsilon_\ell^2) (\lle_{\tau\mu})^2 \right|^2 + \\
	&  \quad + \left| \frac{1}{2} ( (\epsilon_\ell^0)^2 - \epsilon_\ell^2) \lle_{\tau\mu} \right|^2 = 1.0060 \pm 0.0030 ~,
	\label{eq:tauLFU3}
\end{split}\ee
where the first term is due to modification of $\tau_L \to \mu_L \nu_\tau \bar{\nu}_\mu$ while the second is the indistinguishable LFV contribution $\tau_L \to \mu_L \nu_\tau \bar{\nu}_\tau$.
\item Tree-level contributions to LFV $\tau$ decays, both from modified $Z \tau \mu$ couplings and from direct four-lepton operators, which add to the radiative contributions in Eq.~\eqref{eq:LFVtau1}:
\be\begin{split}
	\frac{\mathcal{B}(\tau \to 3 \mu)}{\mathcal{B}(\tau \to \mu \nu \nu)} = & \left[ 2 \left( -2 (g_{\mu L}^Z)^{\rm SM} (\delta g_{\tau\mu L}^Z + \Delta g_{\tau\mu L}^Z ) - ((\epsilon_\ell)^2 + (\epsilon_\ell^0)^2) \lle_{\mu\mu} \lle_{\tau\mu} / 4) \right)^2 + \right. \\
	& \left. + \left( -2 (g_{\mu R}^Z)^{\rm SM} (\delta g_{\tau\mu L}^Z + \Delta g_{\tau\mu L}^Z ) \right)^2 \right]~,
	\label{eq:LFVtau2}
\end{split}\ee
where $\Delta g_{\tau\mu L}^Z = \frac{1}{2} (\epsilon_H^0 \epsilon_\ell^0 + \epsilon_H \epsilon_\ell) \lle_{\tau\mu}$~.
\item Deviation in the electroweak $\hat{T}$ (or $\rho$) parameter \cite{Barbieri:2004qk} due to the custodially-violating operator proportional to $(\epsilon_H^0)^2$:
\be
	\hat{T} = (\epsilon_H^0)^2 \approx  (1 \pm 6) \times 10^{-4}~.
	\label{eq:limTh}
\ee
\end{itemize}


\vspace{0.5cm}

 
\bibliographystyle{JHEP}

{\footnotesize
\bibliography{paper}
}

\end{document}